\documentclass[preprint,12pt]{elsarticle}

\usepackage{graphicx}
\usepackage{amssymb}
 \usepackage{amsthm}

 \usepackage{amsmath}

\usepackage{lineno}

\usepackage{hyperref}

\usepackage{subfig}

\usepackage{multirow}
\usepackage{bigstrut}

\newcommand{\fluence}[2]{$\Phi=#1\times10^{#2}\rm{n_{eq}/cm^2}$}
\newcommand{\fl}[2]{$#1\times10^{#2}\rm{n_{eq}/cm^2}$}

\journal{Nuclear Instruments and Methods A}

\begin{document}

\begin{frontmatter}


\title{Performance of thin planar \textit{n-on-p} silicon pixels after HL-LHC radiation fluences}



\author[1]{A.~DUCOURTHIAL}
\author[1]{M.~BOMBEN\corref{cor1}}
\author[1]{G.~CALDERINI}
\author[1]{R.~CAMACHO}
\author[1]{L.~D'ERAMO}
\author[1]{I.~LUISE}
\author[1]{G.~MARCHIORI}
\author[2,3]{M.~BOSCARDIN}
\author[4]{L.~BOSISIO}
\author[5]{G.~DARBO}
\author[3,6]{G.-F.~DALLA BETTA}
\author[7]{G.~GIACOMINI}
\author[8]{M.~MESCHINI}
\author[9]{A.~MESSINEO}
\author[2,3]{S.~RONCHIN}
\author[2,3]{N.~ZORZI}

\address[1]{LPNHE, Sorbonne Universit\'e, Paris Diderot Sorbonne Paris Cit\'e, CNRS/IN2P3, Paris, France}
\address[2]{Fondazione Bruno Kessler, Centro per i Materiali e i Microsistemi (FBK-CMM)\\   Povo di Trento (TN), Italy}
\address[3]{Trento Institute for Fundamental Physics and Applications (TIFPA INFN), Trento, Italy}
\address[4]{INFN Trieste, Trieste, Italy}
\address[5]{INFN Genova, Genova, Italy}
\address[6]{Universit\`a di Trento, Dipartimento di Ingegneria Industriale, I-38123 Trento, Italy}
\address[7]{Brookhaven National Laboratory, Instrumentation Division 535B, Upton, NY, U.S.A.}
\address[8]{INFN Firenze, Firenze, Italy}
\address[9]{Universita Pisa (IT), INFN Pisa, Pisa, Italy}

\cortext[cor1]{corresponding author}



\begin{abstract}
The tracking detector of ATLAS, one of the experiments at the Large Hadron Collider (LHC),  will  be upgraded  in 2024-2026 to cope with 
the challenging environment conditions of the High Luminosity LHC (HL-LHC). The LPNHE, in collaboration with FBK and INFN, 
has produced 130~$\mu$m thick $n-on-p$ silicon pixel sensors which can withstand the expected large particle fluences at HL-LHC, while delivering data at high rate with  excellent hit efficiency. Such sensors were tested in beam before and after irradiation 
both at CERN-SPS and at DESY, and their performance are presented in this paper. 
Beam test data indicate that these detectors are suited for all the layers  where planar sensors are foreseen in the future ATLAS tracker: 
hit-efficiency is greater than 97\% for fluences of $\Phi \lesssim $\fl{7}{15} and module power consumption is within the specified limits.
Moreover, at a fluence of  \fluence{1.3}{16},  hit-efficiency is still as high as 88\% and charge collection efficiency is about 30\%.
\end{abstract}

\begin{keyword}
Silicon Radiation Detectors \sep LHC \sep HL-LHC \sep Radiation Hardness 


\end{keyword}

\end{frontmatter}


\section{Introduction}
\label{sec:intro}


\noindent CERN plans to upgrade the LHC into a high luminosity machine (High Luminosity LHC, HL-LHC)~\cite{HL_LHC} to expand its physics reach. For this reason the ATLAS detector~\cite{AtlasDetector} will undergo a series of upgrades in the next years. In particular the ATLAS Inner Detector (ID,~\cite{AtlasID1,AtlasID2}) will be replaced starting in 2024 by an all-Silicon system, the ATLAS Inner Tracker (ITk,~\cite{HL-LHC,ITkStripsTDR,ATLASITkPixelTDR}); data taking should resume in 2026. 
The new ATLAS tracking system  will have to assure the same performance as the actual ID but in the much harsher environment of the HL-LHC. The upgraded LHC will deliver 5-7 times larger instantaneous luminosity, which translates into a similar increase of event pile-up, tracks density and radiation fluences/doses with respect to the LHC design values; the goal for HL-LHC is to deliver an integrated 
luminosity of 4000~fb$^{-1}$ by 2037, after 10 years of operation~\cite{ATLASITkPixelTDR}.
The ATLAS ITk will include pixel detectors closest to the interaction point and micro-strip detectors at larger radii. 
The ITk Pixel Detector~\cite{ATLASITkPixelTDR} will comprise 5 barrel layers and multiple rings to cover the very forward region, down to $|\eta|=4.$\footnote{ATLAS uses a right-handed coordinate system with its origin at the nominal interaction 
point (IP) in the centre of the detector and the $z$-axis coinciding with the axis of the beam pipe.  The 
$x$-axis points from the IP towards the centre of the LHC ring, and the $y$-axis points upward. 
Cylindrical coordinates ($r$,$\phi$) are used in the transverse plane, $\phi$ being the azimuthal angle 
around the $z$-axis. The pseudorapidity is defined in terms of the polar angle $\theta$ as $\eta = - \ln 
\tan(\theta/2)$.}

The innermost pixel layers of ATLAS ITk are expected to integrate a radiation fluence $\Phi$ of 1-2$\times$10$^{16} \,\textrm{1 MeV equivalent neutrons}\, (\textrm{n}_{\textrm{eq}})/\textrm{cm}^2$ by the end 
of 2037; this is a factor 4 larger than what the actual ATLAS Insertable B-Layer (IBL~\cite{IBLTDR,IBL_paper}) is expected to have integrated by the end of 
2023. Such a large increase in radiation fluence, with the request of a hit reconstruction efficiency of at least 97\%~\cite{ITkStripsTDR} 
through the whole 
lifetime of the detector, dictates an activity of R\&D toward thin pixel sensors in planar technology, with thicknesses of the order of 100-150~$\mu$m, to mitigate the impact of charge trapping from radiation damage induced defects. As a 
reminder, the ATLAS IBL planar sensors are 200~$\mu$m thick, while outer ATLAS Pixel detector layers feature 250~$\mu$m thick sensors.
The main effect of such large radiation fluences will be the loss of collected signal, which can be as high as 
70\% for the largest HL-LHC fluences even for 200~$\mu$m thick detectors~\cite{IBLTDR,KRAMBERGER2002297}. 

In this paper we report on the beamtest performance of thin $n-on-p$ silicon pixel detectors aimed at the ATLAS ITk pixel innermost layers. 
In Section~\ref{sec:sensors} the characteristics of the joint LPNHE/FBK/INFN pixel production will be presented, together with  details 
of the irradiation campaign some detector modules from that production went through. After having discussed the beamlines used, the tracking telescope, and the data-acquisition, reconstruction and analysis software (Section~\ref{sec:testbeam}), the beamtest results will be 
presented in Section~\ref{sec:results}. Conclusions will be drawn in Section~\ref{sec:conclusions}.


\section{LPNHE/FBK/INFN Thin Sensors Production and Irradiation Campaign}
\label{sec:sensors}

\begin{figure}[h]
\centering
\includegraphics[width=0.65\textwidth]{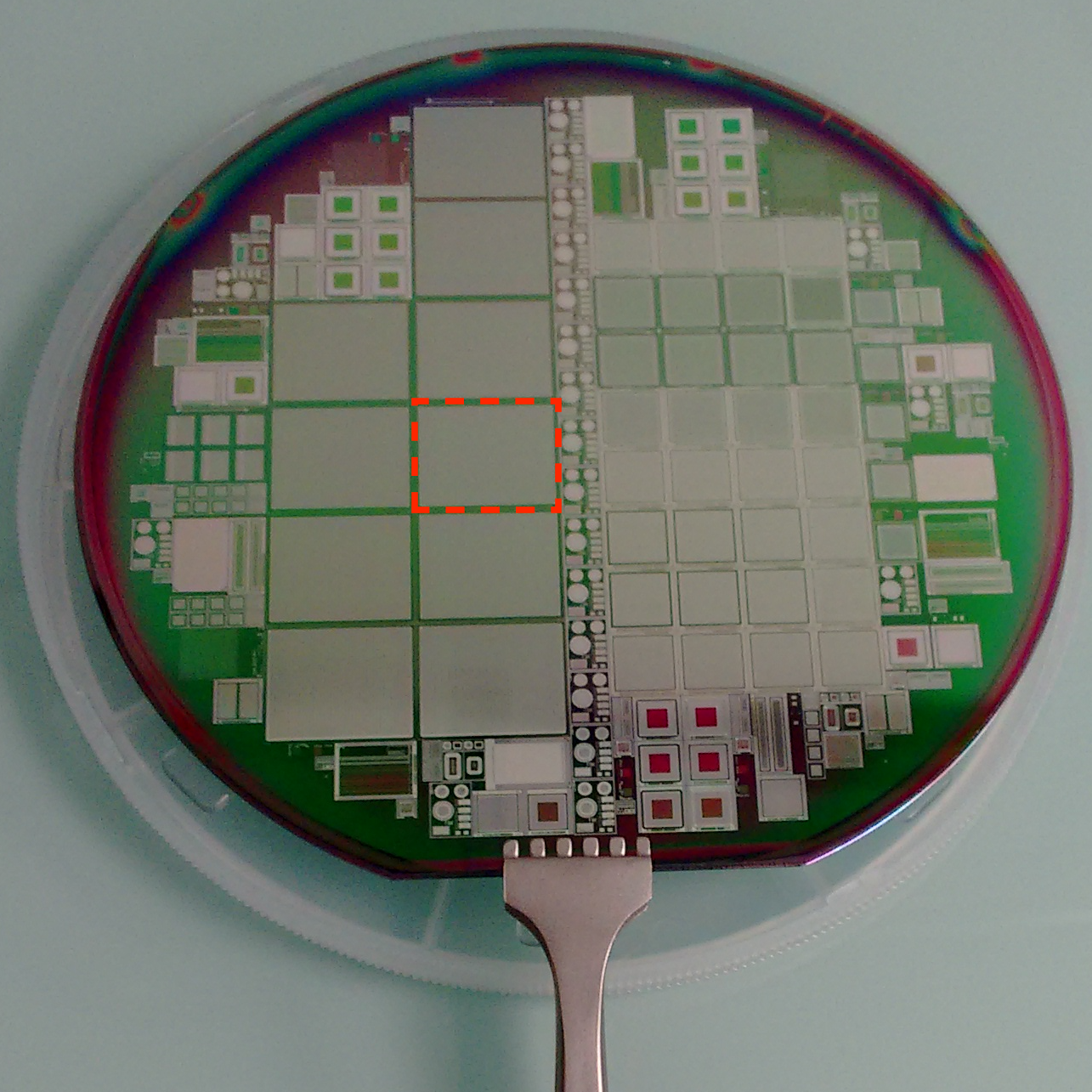}
\caption{\label{fig:wafer.png} Wafer from the $n-on-p$ planar technology production~\cite{DALLABETTA2016388}  whose layout was mainly based on ATLAS FE-I4 and CMS PSI46~\cite{PSI46} designs. The red rectangle
encircles one pixel sensor compatible with the FE-I4 readout chip.}
\end{figure}

\noindent Thin $n-on-p$ planar pixel sensors have been realised at FBK\footnote{FBK-CMM (Trento, Italy): \url{http://cmm.fbk.eu/}} on high resistivity p-type 150~mm (6'')
wafers within
the  framework of the INFN Phase-2 program~\cite{DALLABETTA2016388}.
Si-Si Direct Wafer Bonded (DWB) wafers were chosen to fabricate pixel detectors;  they  are obtained by bonding together two 
different 
wafers: a high-resistivity (HR) Float Zone (FZ) sensor wafer and a low-resistivity (LR) Czochralski handle wafer. The FZ wafer is thinned to the 
desired thickness value, so as to obtain a wafer with a thin active layer plus a relatively thick mechanical support layer. P-type wafers of 
two 
different active depths (100 and 130~$\mu$m) with 500~$\mu$m  thick handle wafers were used.
The wafer layout included sensors compatible with one and two FE-I4 chip~\cite{FEI4} modules (one chip module
 surface $\sim$ 20$\times$16.8~mm$^2$), with pixel cell size 
of 50$\times$250~$\mu$m$^2$. In Figure~\ref{fig:wafer.png} a picture of one wafer from this production is shown.
The ITk Pixel Detector modules will be equipped with a completely new readout chip; a first prototype, the 
RD53A~\cite{RD53}, was recently produced. The pixel cell size is of 50$\times$50~$\mu$m$^2$; it can be used to 
readout a sensor with pixel cells of the same size but also the 25$\times$100~$\mu$m$^2$ option is currently being explored.

\begin{figure}[h]
\centering
\includegraphics[width=0.49\textwidth]{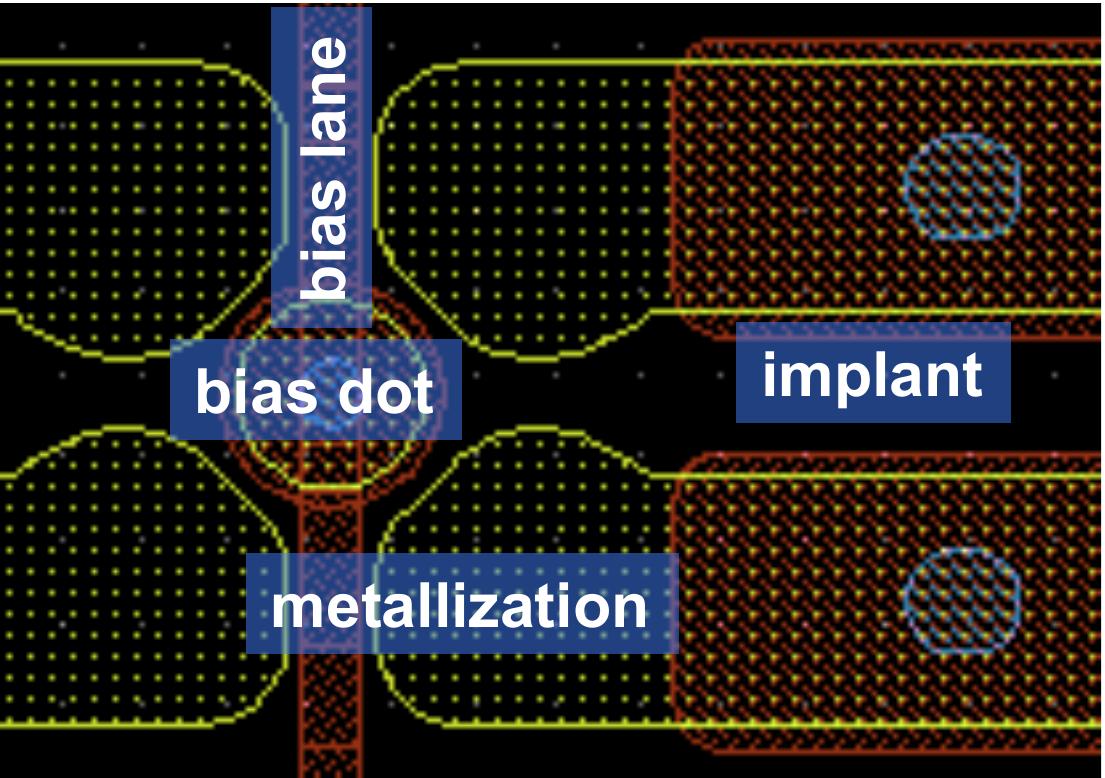}
\includegraphics[width=0.49\textwidth]{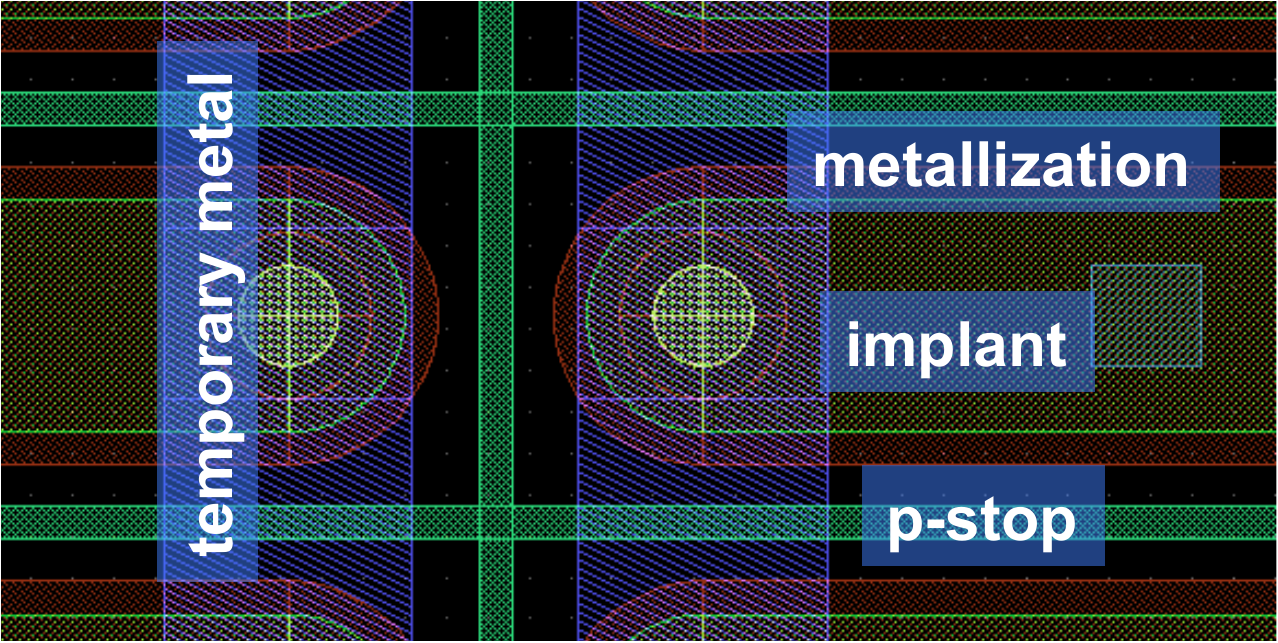}
\caption{\label{fig:biasdot}(left) Schematic view of parts of pixel cells presented in this article. The salient parts are indicated. 
(right) The scheme of pixels cell from a previous production~\cite{1748-0221-12-05-P05006} is reported, where temporary metal lanes 
where used for biasing before bump-bonding.}
\end{figure}

Permanent biasing structures were implemented on the pixel sensors, consisting of a small circular $n^+$  implant (\textit{bias dot}) in the 
corner between four neighbouring pixel cells; all bias dots were shorted together through a metal line (\textit{bias lane}). Thanks to these 
structures, by exploiting the punch-through mechanism,  the pixels could be tested electrically before  bump-bonding to the readout chips. In Figure~\ref{fig:biasdot} a scheme of the pixels cells.
The diameter of the implant of the bias dot is of 27~$\mu$m; the metallization diameter is 4~$\mu$m smaller. The pixel 
implant is 37~$\mu$m high and about 200~$\mu$m long; the metallization of the pixel cell is about 3~$\mu$m shorter on both 
sides.
As a comparison in Figure~\ref{fig:biasdot} 
a scheme of pixel cells from a previous production~\cite{1748-0221-12-05-P05006} 
is shown, where the temporary metal solution~\cite{bib:metal} was used to bias the sensor before bump-bonding. In this case 
the pixel implant is 27~$\mu$m high and about 220~$\mu$m wide. The pixel metallization is 4~$\mu$m wider than the 
implant.

\subsection{Pixel module irradiation}
\label{sec:irradiation} 
\noindent Radiation hardness  was tested by  measuring the performance of irradiated pixel sensors connected to   FE-I4 readout chips. Two 
sensors, named {\it W80} and {\it W30}, were taken from two different sensor wafers,
with thickness of 130 (100)~$\mu$m for W80 (W30); the sensors 
 had different number of guard rings (GRs),  2 and 5, respectively. Sensors details are summarised in Table~\ref{tab:W30W80Irr}.  In both detector assemblies the 500~$\mu$m  thick handle wafer was not thinned.
A Benzo-­Cyclo-­Butene (BCB)   layer was deposited on  the sensors for spark    protection (for more details~\cite{Stefano,UNNO201372}). Each sensor was  bump bonded to an FE-I4 chip at 
IZM, Berlin\footnote{Fraunhofer-Institut f\"ur Zuverl\"assigkeit und Microintegration: \url{https://www.izm.fraunhofer.de/en.html}}.

The irradiations of W80 and W30 were carried at room temperature at the CERN IRRAD facility\footnote{\url{http://ps-irrad.web.cern.ch/}} using a 24~GeV/c proton beam. For both pixel modules the irradiation was staged in two steps.
Table~\ref{tab:W30W80Irr} gives the details of the irradiation program for the two modules tested from
that production, W80 and W30; their characteristics are reported too. It has to be noted that the IRRAD beam profile is gaussian 
with FWHM ranging from 12 to 20~mm. \begin{table}[h]
\caption{\label{tab:W30W80Irr}Irradiation program for the two FE-I4 pixel modules W80 and W30. For both pixel modules 
the irradiation was staged in two steps; the average fluence $\phi$ is reported for each step. The average cumulative 
fluence $\Phi$ after the second step is reported too.}
\centering
\begin{tabular}{cccc}
\hline
Module name & Beam spot size & $<$Fluence$>$ $\phi$& $<$Cumulative fluence$>$ $\Phi$ \\
(thickness [$\mu$m], \# of GRs) & (FWHM - [mm$^2$]) &  [10$^{15}$ n$_\text{eq}/\text{cm}^2$]  &  [10$^{15}$ n$_\text{eq}/\text{cm}^2$] \\
\hline
\hline
W80 (130, 2) & 20$\times$20 & 3 & 3\\
\hline
W30 (100, 5) & 12$\times$12 & 4 & 4\\
\hline
W80 (130, 2) & 20$\times$20 & 7 & 10 \\
\hline
W30 (100, 5) & 20$\times$20 & 7 & 11\\
\hline
\end{tabular}
\end{table}

In summary W80 (W30) received during the first irradiation experiment an average fluence  $\phi$ of about 
3~(4)~$\times$10$^{15}$~n$_\text{eq}/\text{cm}^2$ and of 7~(7)$\times$10$^{15}$~n$_\text{eq}/\text{cm}^2$ 
during the second irradiation experiment; after the second irradiation experiment the average cumulative fluence  $\Phi$ 
received by W80 (W30) was of about  1.0~(1.1)$\times$10$^{16}$~n$_\text{eq}/\text{cm}^2$.

\begin{figure}[h]
\centering
\includegraphics[width=0.49\textwidth]{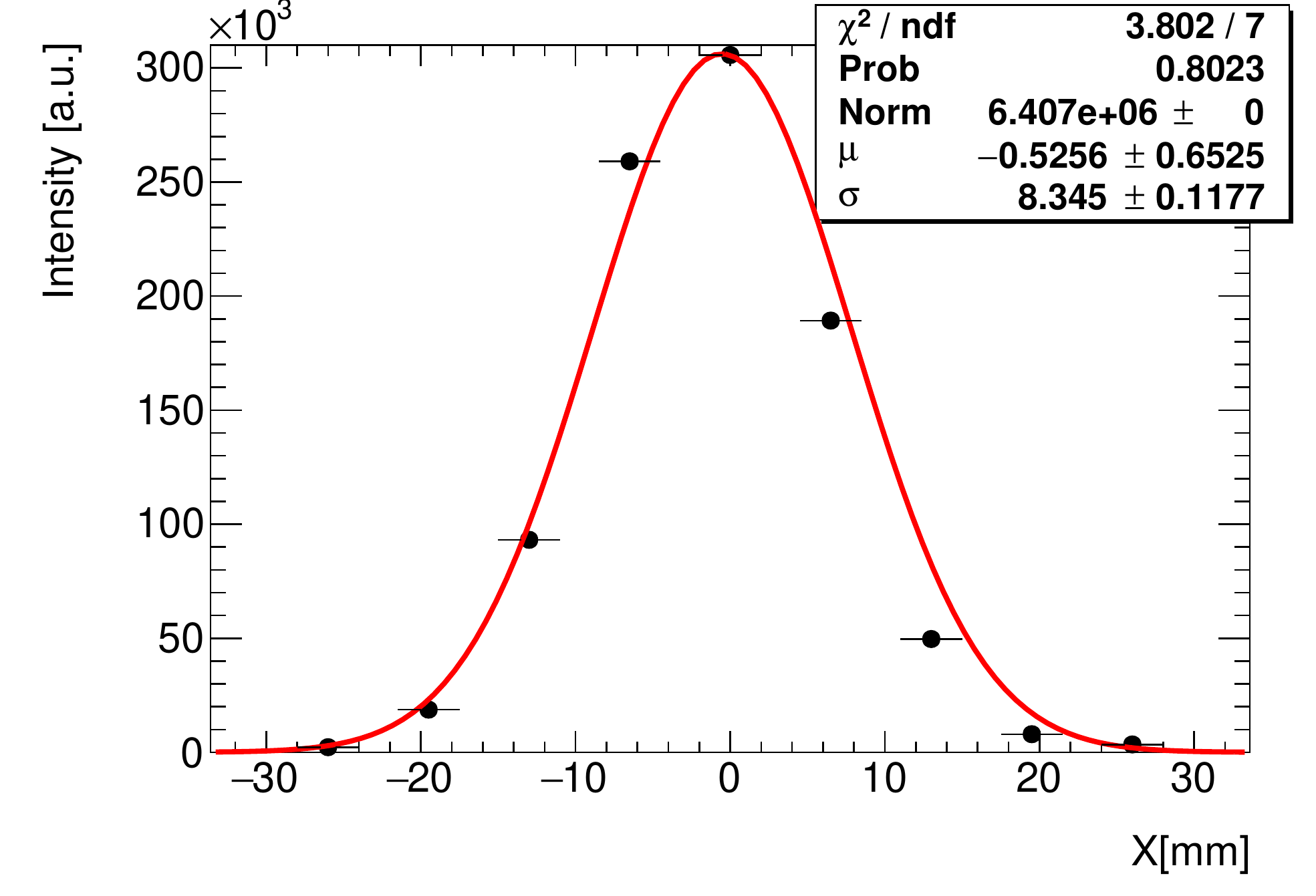}
\includegraphics[width=0.49\textwidth]{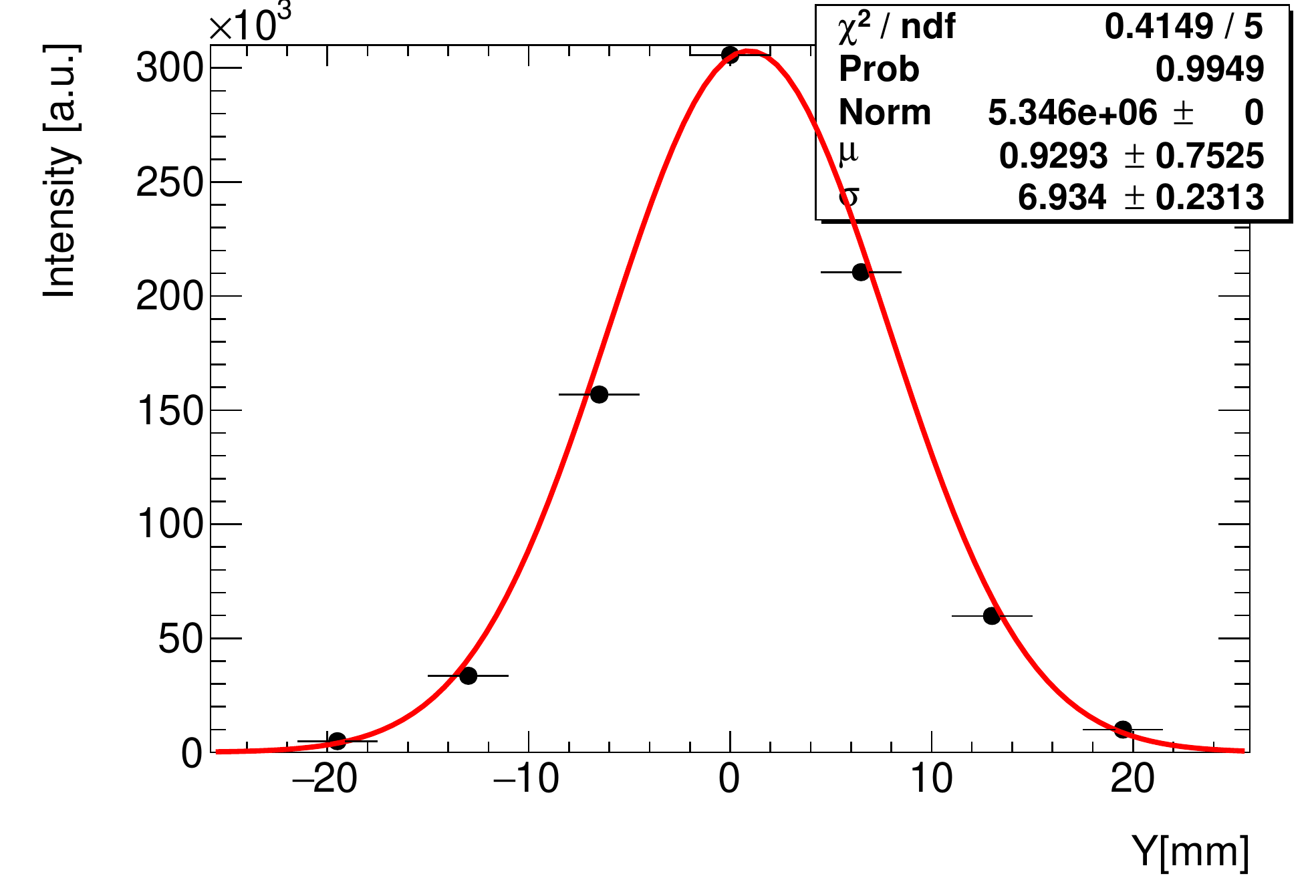}
\caption{\label{fig:W80_irr_1D} Projections of the proton beam profile used at CERN IRRAD to irradiate the W80 module. Left: horizontal 
direction; right: vertical direction. A gaussian fit is superimposed. The (0,0) position correspond to the nominal beam center.}
\end{figure}

 At the IRRAD facility several beam position monitors (BPMs), which register the 
beam intensity during the irradiation along the horizontal and vertical direction orthogonal to the beam, allow reconstructing the beam 
profile. 
The accuracy in the position determination is of the order of 
2~mm, which includes the different sources of misalignment. Figure~\ref{fig:W80_irr_1D} shows 
the beam profile intensity projected along the 
horizontal and vertical direction; the two projections have been fitted with a gaussian to determine the center position and the beam 
widths.
It can be seen that the center vertical position is not compatible with $y=0$; this has been confirmed by the IRRAD facility managers.

\begin{figure}[h]
\centering
\includegraphics[width=0.65\textwidth]{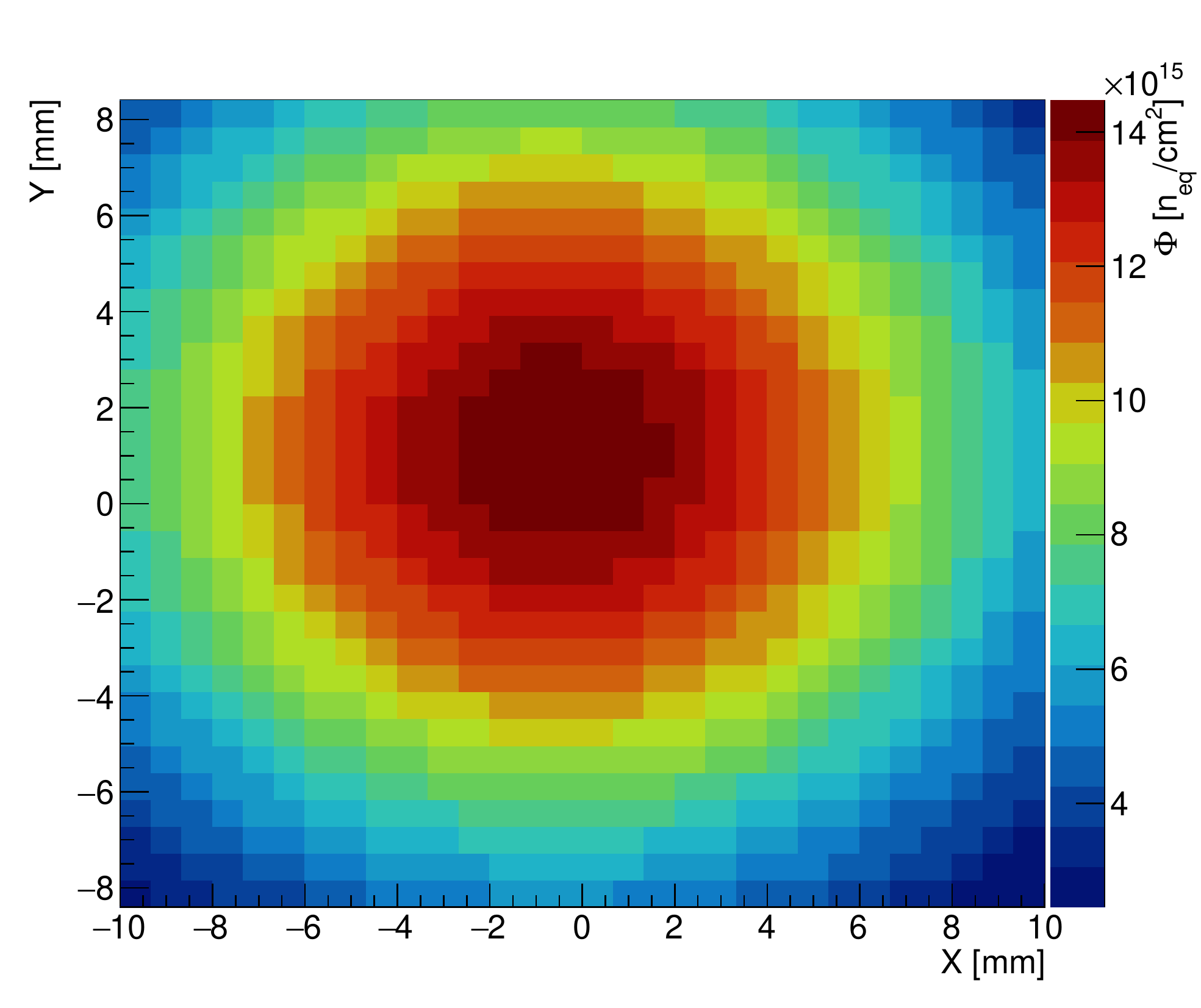}
\caption{\label{fig:W80_irr_2D} Fluence profile for W80 module after the the second irradiation step. The area covered by the figure corresponds 
to the surface of the W80 pixel module.}
\end{figure}

The dosimetry information made possible to estimate the total delivered proton fluence, transformed then into n$_\text{eq}/\text{cm}^2$ using an hardness factor~$\kappa=0.59$, with an uncertainty of about 10\%.
In Figure~\ref{fig:W80_irr_2D} the fluence profile after the second irradiation step is reported for the W80 module.
It can be seen that the highest fluence is  about $\Phi=1.4\times10^{16}$~n$_\text{eq}/\text{cm}^2$, while at the detector periphery the 
fluence  is as low as $\Phi=3.5\times10^{15}$~n$_\text{eq}/\text{cm}^2$.
Thanks to the high segmentation of the pixel detector modules it was then possible to probe several fluences over a large range of values 
with just one pixel detector.

\subsection{Electrical Performance after Irradiation}
\begin{figure}[h]
    \centering
\includegraphics[width=\textwidth]{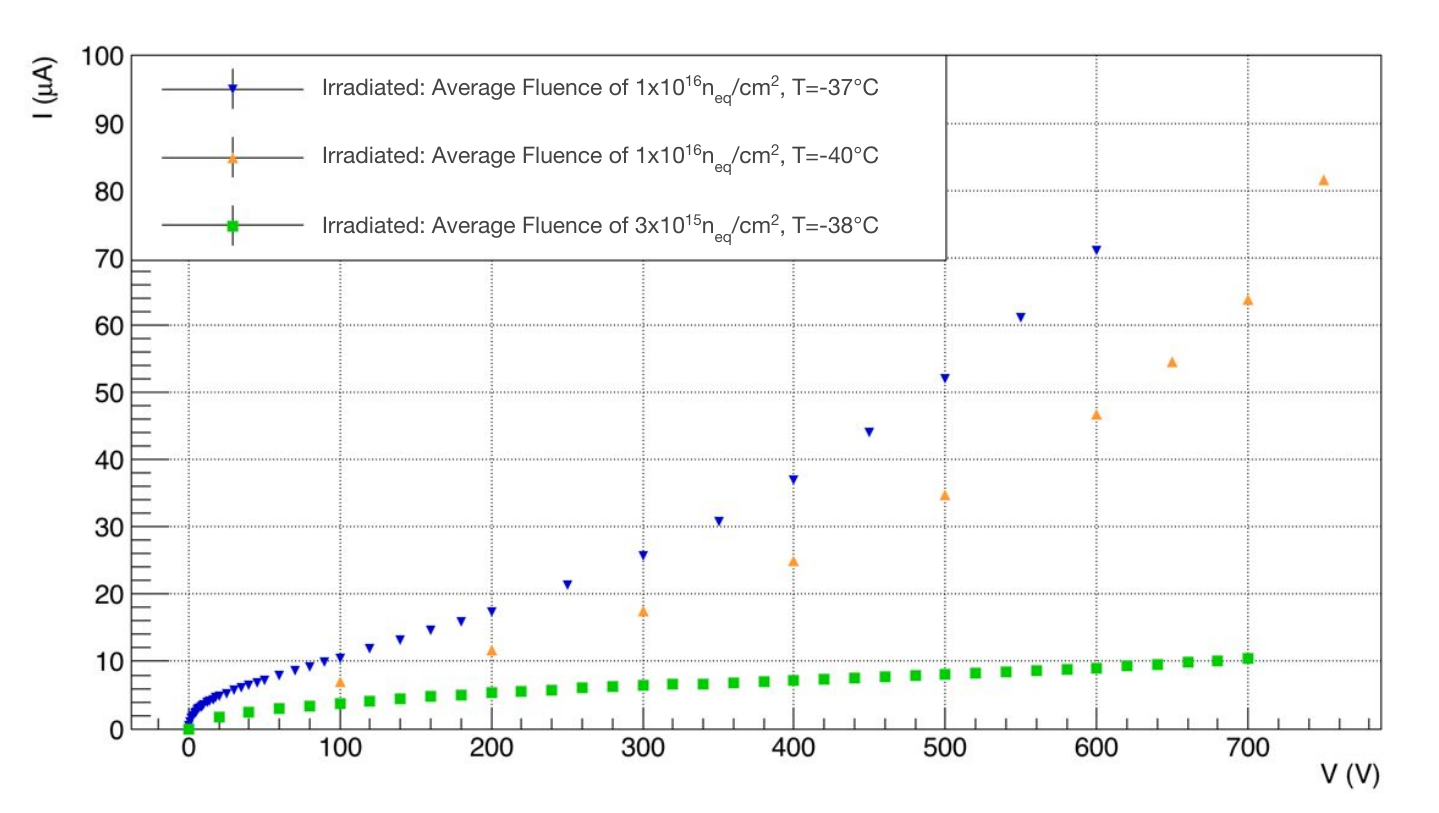} 
\caption{\label{fig:IV}Current-Voltage curves of W80 sensor after a fluence 
 of $3 \times 10^{15}\, \rm{n_{eq}/cm^2}$ (green markers) 
 and after a cumulative fluence of $1 \times 10^{16}\, \rm{n_{eq}/cm^2}$ (blue and yellow markers). The temperature at which the measurements were taken is indicated in the legend.}
\end{figure}

\noindent In what follows the pixel sensor leakage current and power consumption performance are presented  as a 
function of the bias voltage. The average cumulative fluence $\Phi$ from Table~\ref{tab:W30W80Irr} is 
used as a measure of the irradiation fluence received by the module itself\footnote{The leakage current per pixel cell was not 
recorded; only the leakage current of the whole module is available}.
The leakage current as a function of the bias voltage for different fluences and at different temperatures is reported in Figure~\ref{fig:IV} for  the W80 pixel module after irradiation. 

The trend of the leakage as of function of  bias voltage for the largest fluence seems compatible with a 	convex function, while 
the one for the lowest fluence looks more like a concave function. One possible explanation for the trend at large fluence is the 
onset of impact ionisation, leading to current multiplication.

The increase of leakage current  $I$  with the fluence $\Phi$ is expected to be linear: $ \Delta I = \alpha V \Phi$, where $\alpha \sim 4\times 10^{-17} \text{A/cm}$ is the so-called {\it current related damage rate}~\cite{moll-thesis} and $V$ is the volume of the sensor. The data reported in Figure~\ref{fig:IV} were used to extract the current related damage rate $\alpha$ at $V_{bias}~=~ 600$~V after rescaling the current to $t=20^{\circ}$~C. The leakage current was rescaled according to the formula of 
Ref.~\cite{Chilingarov_tscale} ($I(T)\sim T^{2}\exp(-E_{eff}/(2k_{B}T)$) with two $E_{eff}$ values, 
{\it i.e.} 1.12 and 1.21~eV; two $E_{eff}$ values were considered because recent measurements\footnote{MPG ATLAS group, private communication} favour a value for $E_{eff}$ close to the energy gap $E_g$. 

\begin{table}[htbp!]
\centering
\caption{Current related damage rate $\alpha$ values for W80 sensors at  $V_{bias} = 600$~V for two fluences and different temperatures. The uncertainty on the values is due to the temperature uncertainty of 1$^{\circ}$~C. }
\label{alpha_w80}
\def\arraystretch{1.5}
\begin{tabular}{cc|c|c|}
\cline{3-4}
                                                       &                  & \multicolumn{2}{c|}{\begin{tabular}[c]{@{}c@{}}$\alpha$\\ ($10^{-17}$A/cm)\end{tabular}} \\ \cline{3-4} 
                                                       &                  & $E_{eff}=1.12$~eV                           & $E_{eff}=1.21$~eV                           \\ \hline
\multicolumn{1}{|c}{$\Phi=3\times10^{15}\rm{n_{eq}/cm^2}$,} & $t$ = -38$^{\circ}$~C & $2.6\pm0.4$                              & $4.0\pm0.6$                              \\ \hline
\multicolumn{1}{|c}{$\Phi=1\times10^{16}\rm{n_{eq}/cm^2}$,} & $t$ = -40$^{\circ}$~C & $5.2\pm0.7$                              & $8.2\pm1.2$                               \\ \hline
\multicolumn{1}{|c}{$\Phi=1\times10^{16}\rm{n_{eq}/cm^2}$,} & $t$ = -37$^{\circ}$~C & $5.3\pm0.7$                              & $8.2\pm1.2$                               \\ \hline
\end{tabular}
\end{table}

As it can be seen in Table~\ref{alpha_w80}, for all fluences, temperatures and $E_{eff}$ values the current related damage rate $\alpha$ values are in the correct 
ballpark. The two results for the largest fluence at two slightly different temperatures are in good agreement, which 
indicates the reproducibility of the method.
\\
From  Table~\ref{alpha_w80} it can be seen that the value of $\alpha$ for the largest fluence can be as high as twice the ``standard 
value'' ($4\times 10^{-17} \text{A/cm}$ ). This result, together with the fact that the leakage current curve is compatible with a convex 
function, might be a strong argument in favour of the onset of impact ionisation. It has  to be anyhow noted that  pixel modules did not 
undergo the ``standard'' annealing of 80 minutes at 60$^{\circ}$~C  (they were kept cold most of the time after irradiation and they 
were always measured at below 0$^{\circ}$~C temperature\footnote{They spent some unknown amount of time at room temperature after irradiation at IRRAD facility}); hence it cannot be ruled out a contribution to the leakage current to 
non-annealed defects~\cite{moll-thesis}.

\begin{figure}[h]
    \centering
\includegraphics[width=\textwidth]{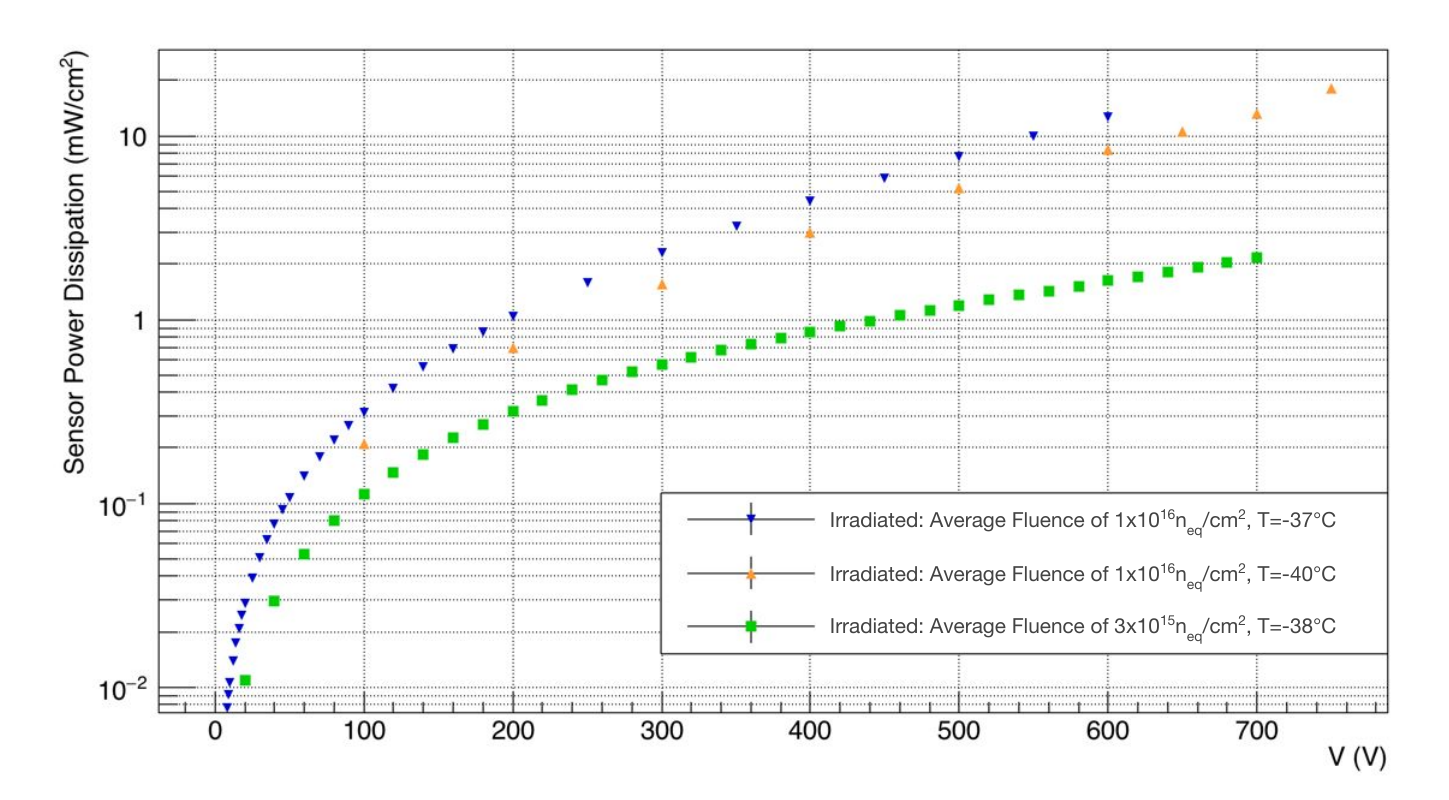} 
\caption{\label{fig:powdiss}Sensor power dissipation curves of W80 sensor after a fluence of \fluence{3}{15} (green markers) and after a fluence of \fluence{1}{16} (blue and yellow markers). The temperature is indicated in the legend.}
\end{figure}

In Figure~\ref{fig:powdiss} the power dissipation per unit area of the W80 detector after irradiation is reported as a function of  bias voltage, for different fluences and temperatures; the power dissipation was computed  as the product of leakage current times bias voltage. 

At a bias voltage of 600~V the power dissipation per unit area
 is about 6~mW/cm$^2$ after a fluence of $3 \times 10^{15}\, \rm{n_{eq}/cm^2}$ 
when scaled to 
$t = -25^{\circ}$~C (the expected pixel detector temperature at ITk~\cite{ATLASITkPixelTDR}) using the same formula used for the leakage current; this power dissipation  is just below 
the specification for ITk pixels at $\Phi=2 \times 10^{15}\, \rm{n_{eq}/cm^2}$ (6.4 mW/cm$^2$~\cite{ATLASITkPixelTDR}). At 
$\Phi=1 \times 10^{16}\, \rm{n_{eq}/cm^2}$ the power dissipation  is about 40 mW/cm$^2$; 
this value is comparable to what has been reported in~\cite{Savic:2016xwn} for 100~$\mu$m thick pixel detectors aimed at ITk.

\section{Testbeam Data Taking}
\label{sec:testbeam}

\noindent Measurements reported here were carried out in 2017 and 2018 at the DESY beam test facility~\footnote{http://testbeam.desy.de/} and  at the CERN-SPS experimental area~\footnote{http://sba.web.cern.ch/sba/}, using copies of the EUDET/AIDA telescope~\cite{Jansen2016}. 
  At DESY, 4~GeV/c electrons from an almost continuous beam were used. At CERN, 120~GeV/c positive pions were used.  The time structure of the 
beam was organised in spills within a super cycle of several tens of 
seconds. 
In what follows a summary of the  experimental conditions will be given;  more details of the testbeam facilities, tracking telescopes, data acquisition system, reconstruction and analysis software used can be found 
in Refs.~\cite{1748-0221-12-05-P05006}~and~\cite{1748-0221-7-10-P10028}.

The devices under test (DUTs) were mounted between the two arms of the tracking telescope. Each telescope arm comprised three 
Mimosa26~\cite{mimosa26} sensing planes. Each Mimosa26 sensor matrix is composed by 576$\times$1152 pixels of 18.4~$\mu$m pitch. 
During the measurements the DUTs were housed in a cooling box which assured the DUTs were at controlled temperature
 (down to -50$^{\circ}$~C) and protected from light. 
  Data acquisition was triggered by the coincidence signal of two plastic scintillators, whose overlap area was
   about 1~cm$^2$.

Prior to  data taking, the DUTs were carefully tuned to choose a threshold value and the correspondence between the Time-over-Threshold (ToT)~\cite{FEI4} value and input charge is calibrated.  In our DUTs,
the signal generated by a MIP\footnote{Minimum Ionising Particle} traversing the sensor is digitized into a 4 bit ToT register. The threshold is chosen to assure high signal 
efficiency while minimising the noise. For our thin sensors  (100-130~$\mu$m thickness) thresholds ranging between 700~e and 1200~e were chosen. Once 
a threshold is selected, a ToT-to-charge calibration has to be performed: a  ToT value will be related to a corresponding amount of charge 
induced on the electrodes. Usually this calibration is tuned to match  ToT values in the middle of the dynamic range of the 4-bit 
register (5-8) to the expected signal of a MIP in the sensor. 
For example, in 130~$\mu$m thick sensors, a MIP is expected to generate about 10~ke, hence a typical ToT tuning for such a detector would be 5 ToT for a signal of 10~ke. After irradiation, as 
charge carriers are trapped, the signal amplitude decreases and a lower charge per ToT unit  is better suited, such as 6 ToT for a signal of 6~ke.

\section{Testbeam results}
\label{sec:results}
\noindent In this section the testbeam results will be presented after a  discussion of the corrections applied to the fluence map~(Sections~\ref{sec:corrections}~and~\ref{sec:OtherSyst}) presented in 
Section~\ref{sec:irradiation}. Data presented in this section 
were reconstructed using EUtelescope~\cite{eutelescope} and analysed using the TBmon2~\cite{tbmon2} framework.

\subsection{Corrections to Fluence Map}
\label{sec:corrections} 

\noindent The fluence maps of W80 have been presented in Section ~\ref{sec:irradiation}. From dosimetry results, the fluence beam profile (see 
Figure~\ref{fig:W80_irr_2D}) can be modeled by 2D gaussians, with a 2~mm uncertainty on the position in both X and Y directions.
To further constrain the fluence peak position and reduce the uncertainties on this position, the mean cluster ToT distribution across the 
sensor was used. As the charge trapping effect increases with the fluence, the collected charge and consequently the ToT are also reduced. 
Hence the position of the minimum of the mean  ToT distribution of a cluster is a valid indicator of the fluence peak position. 
For this purpose, various configurations in terms of threshold, ToT configuration and bias voltage have been investigated, as
reported in Table~\ref{tab:configurations}

\begin{table}[h]
\centering
\caption{\label{tab:configurations}Summary of the various configurations investigated for W80 after the second irradiation step. 
The ToT tuning describes the ToT value corresponding to a  charge collected by the electrode. The charge target value
(in units of thousand of electrons) is indicated by $i$. }
\begin{tabular}{|c|c|c|c|c|c|c|}
\hline
Threshold [electrons] & 850 & 850 & 850 & 1000 & 1000 & 1200 \\
\hline
ToT at $i$ ke- & 8 at 4 & 8 at 4 & 8 at 4 & 8 at 4 & 6 at 4 & 6 at 4 \\
\hline
Bias Voltage [V] & 600 & 500 & 400 & 600 & 600 & 600 \\
\hline
\end{tabular}
\end{table}

\begin{figure}[h]
\centering
\includegraphics[width=\textwidth]{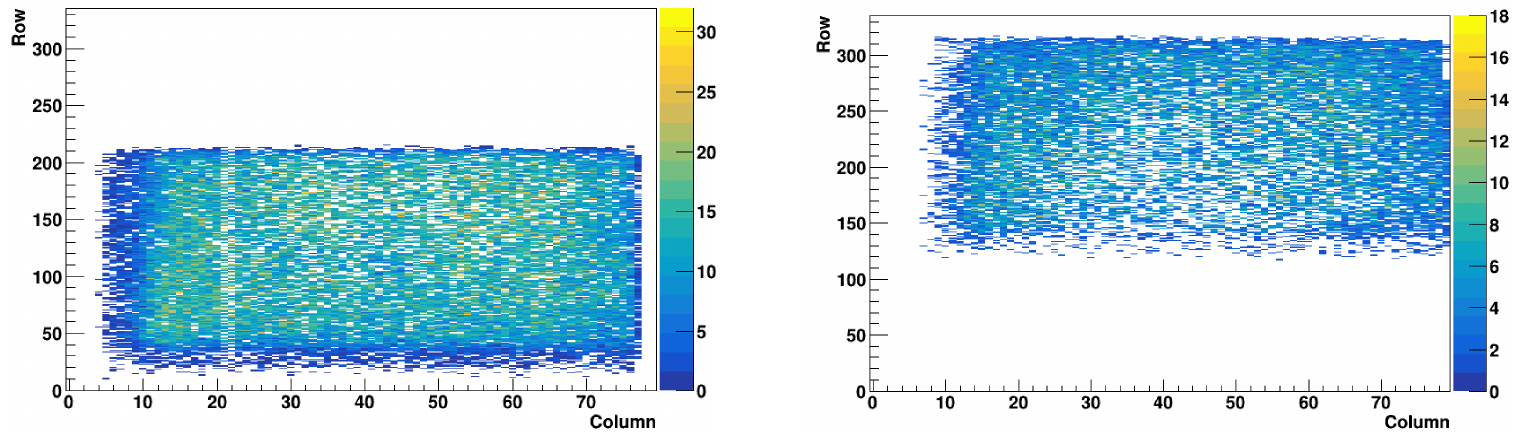}
\caption{\label{fig:ROIs}Hit map of W80. The Down and Up ROI are visible respectively in the left and right plots.}
\end{figure}

The search of the irradiation beam profile position has been performed with data taken at DESY where the beam profile was wider than what it was at CERN-SPS.
During this testbeam two ROI (Regions Of Interest) have been considered which area were covered by the beam with a sufficient amount of statistics (the edges of the beam profiles where the statistics is too limited are not considered). The ROI covering the lower part of the sensor will be referred as ``Down'' position in the following; the other covering the upper-medium part of the sensor will be referred as ``Up'' position in the following\footnote{The choice of the regions was dictated 
by the triggering scintillators coverage}. Both ROIs are presented in Figure~\ref{fig:ROIs}

\begin{figure}[h]
\centering
\includegraphics[width=\textwidth]{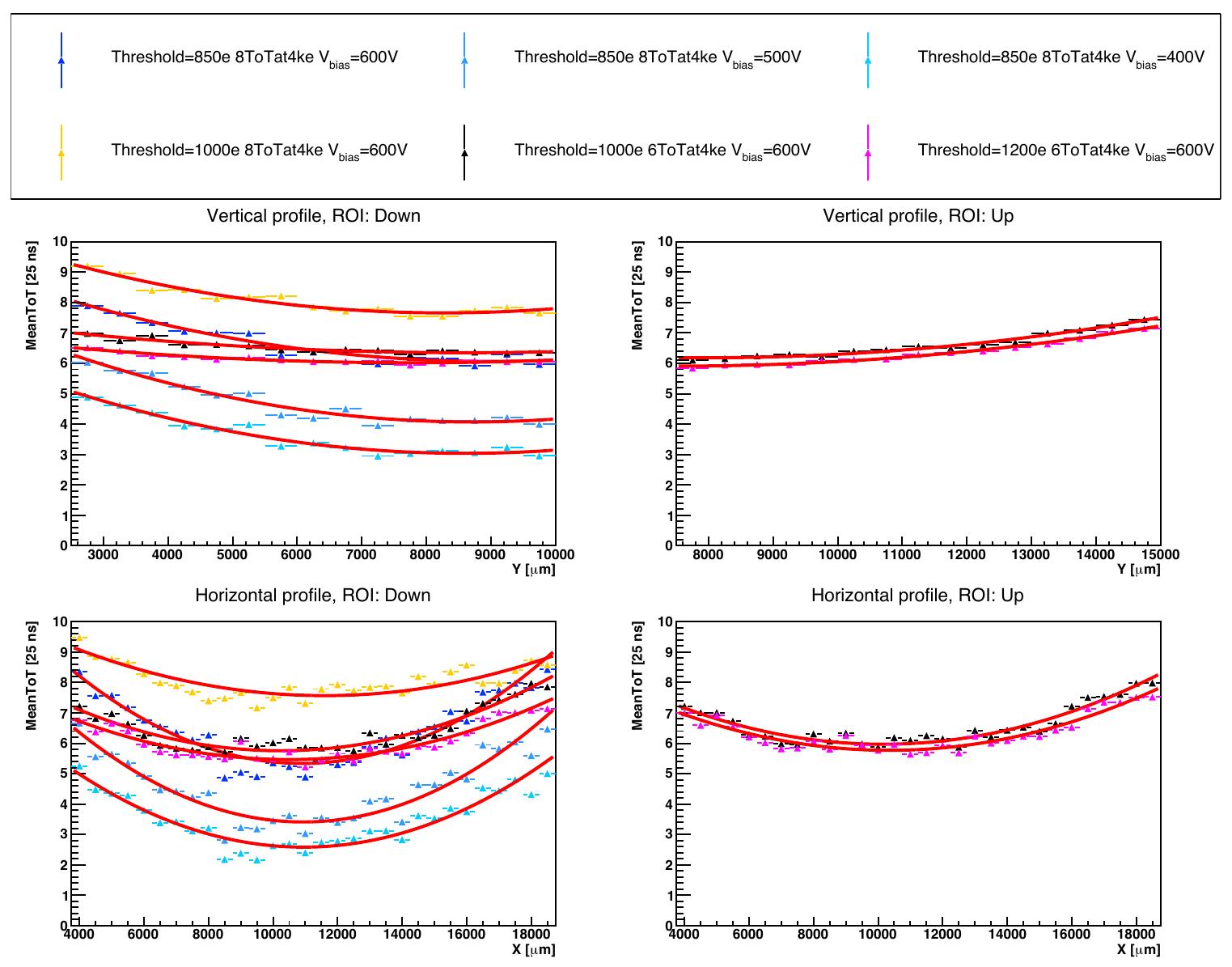}
\caption{\label{fig:ToT_profiles}Average ToT profiles for the two ROI. The threshold, ToT tuning and bias voltage are indicated in the legend 
box. Each distribution is fitted with a polynomial of degree two (red lines).}
\end{figure}

\noindent Two profiles have been created, an horizontal profile which averages all the mean ToT values of each pixel along the vertical axis 
in the ROI and a vertical profile which averages all the mean ToT values of each pixel along the horizontal axis in the ROI.
To obtain the value of the peak fluence position, the average ToT profiles have been created for all the configurations from 
Table~\ref{tab:configurations} and they are presented in Figure~\ref{fig:ToT_profiles}.

Each distribution is fitted with a 2$^{\rm nd}$ degree polynomial and the minimum of the distribution is extracted from the fit. The mean of the average 
ToT minimum position value, which corresponds to the fluence peak position, is obtained by averaging the extracted values of all 
configurations. The fluence peak position estimated in this way is shifted by 1.2~mm in X and by 1~mm in Y with 
respect to the position presented in Section~\ref{sec:irradiation}; the shift is within 
the 2~mm uncertainty on the position indicated by the CERN IRRAD facility.



\begin{figure}[h]
\centering
\includegraphics[width=\textwidth]{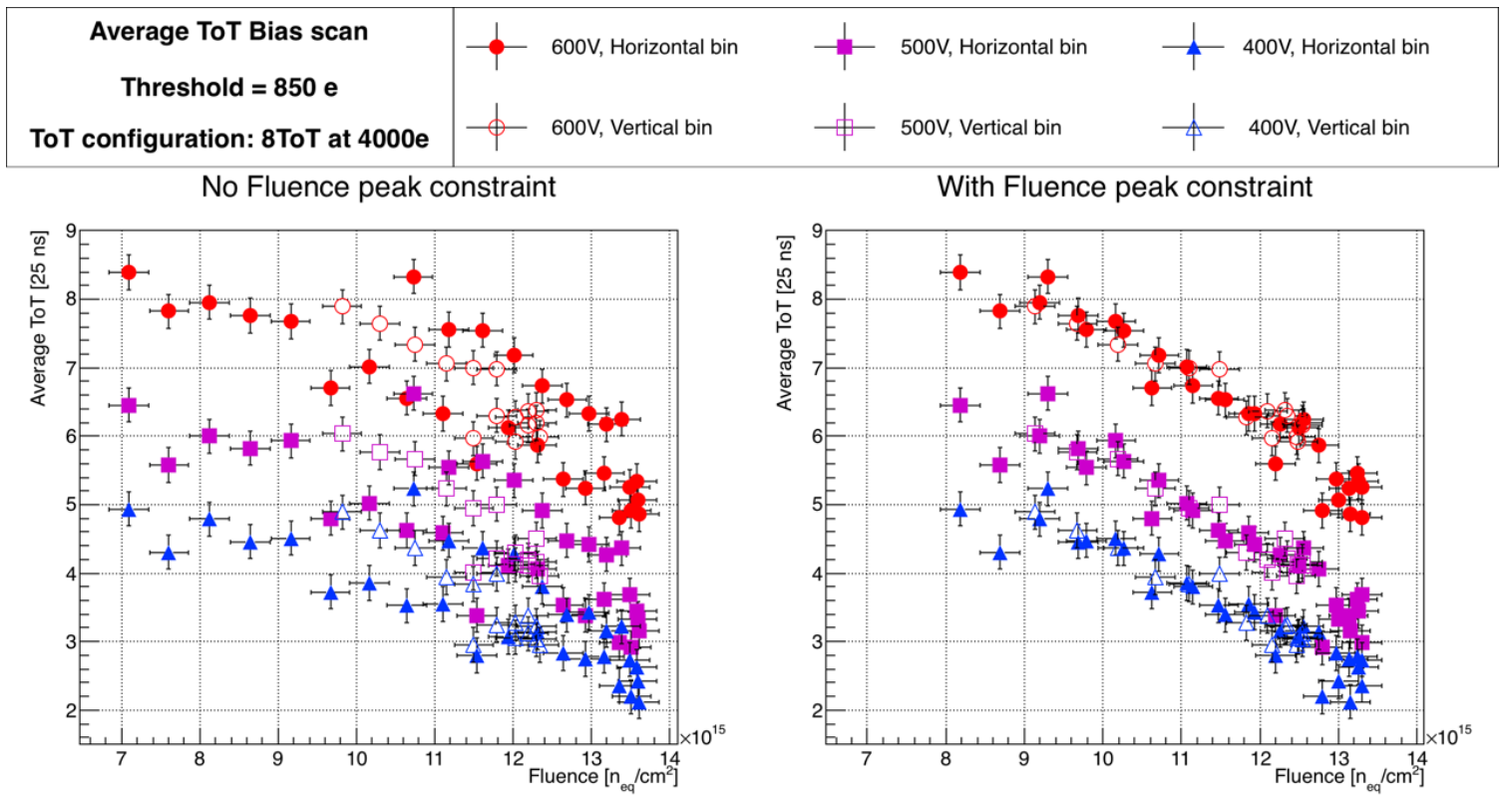}
\caption{\label{fig:modifications}Average ToT distribution vs fluence for three different bias voltages. The left/right plot is without/with fluence 
peak constraint. The horizontal/vertical bin label in the legend means that the fluence and average ToT have been extracted from an 
horizontal/vertical profile of the Down ROI.}
\end{figure}

The effect of the modification of the peak fluence is presented in Figure~\ref{fig:modifications} which shows the average ToT vs fluence for 3 
different bias voltages without (left plot) and with (right plot) fluence peak constraint. The constraint of the fluence peak (right plot) results in 
less dispersion in the average ToT values for the same fluence.

 The uncertainty on the fluence in Figure~\ref{fig:modifications} is set to $0.5\times10^{15} \text{n}_\text{eq}/\text{cm}^2$, which corresponds to the
  variation of the fluence at 2~mm of the peak value. It accounts for the uncertainty on the peak fluence and on the fluence profile modelling 
  approximation.

 \subsection{Other Systematic Effects on Fluence Determination}
 \label{sec:OtherSyst}
 
\begin{figure}[h]
\centering
 \subfloat[MeanToT map of reference DUT]{\includegraphics[width=0.545\textwidth]{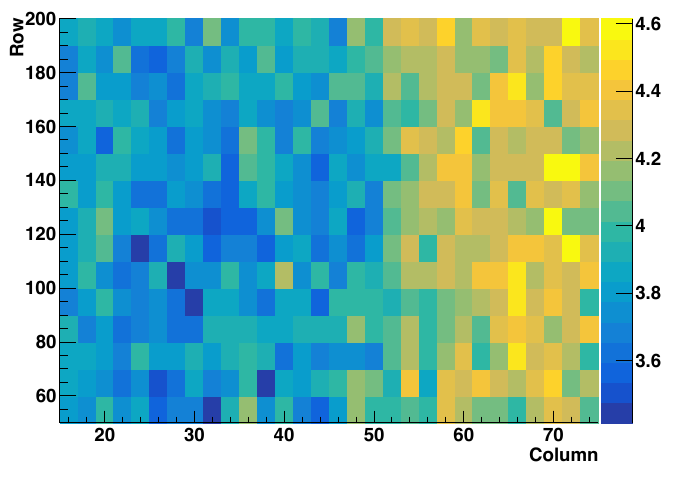}}
  \subfloat[Horizontal profile]{\includegraphics[width=0.455\textwidth]{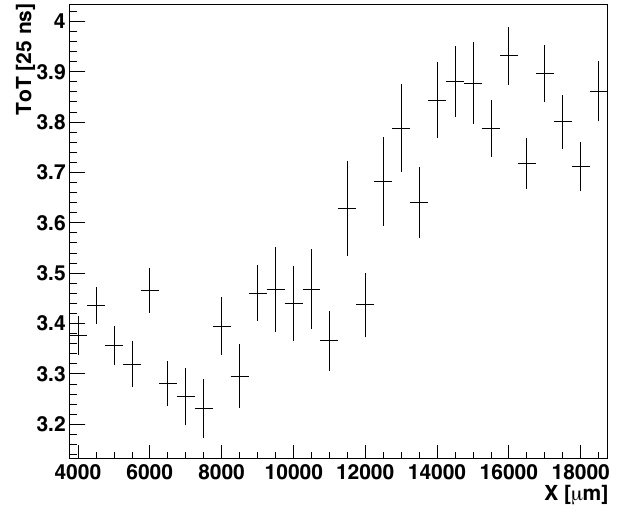}}
  \caption{\label{fig:ToTDispersion}Cluster ToT map (a) and ToT horizontal profiles for the reference DUT (un- irradiated). 
  The DUT tuning was 4~ToT at 8~ke-.}
 \end{figure}

\noindent Even if the constraint on the fluence peak shows better results in terms of dispersion, some other effects have to be accounted for. 
The average ToT distribution is highly sensitive to the chosen threshold and ToT value, as well as to the bias voltage. 
As several tunings have been investigated and give consistent results with respect to one another, the peak fluence position is assumed 
rather independent of the tuning.
Another effect is the non uniformity of the threshold across the different pixels observed in the FEI4 chip. This has an impact on the average ToT, as presented in 
Figure~\ref{fig:ToTDispersion}, which shows the average ToT distribution on the un-irradiated reference DUT used in the testbeam where 
the previous data were extracted.
 A variation of the average ToT of about 0.5 ToT can be seen across the sensor even if this sensor was un-irradiated.  The difference in ToT occurs mainly horizontally. This ToT variation can be explained by a drift in the threshold already seen by other users 
 of FEI4 modules\footnote{MPG ATLAS group, private communication}. To take into account this effect, which is difficult to quantify from one 
 chip to another, a conservative systematic uncertainty of 0.5 ToT was assigned to the ToT value .
 
\subsection{Hit Efficiency}
\label{sec:hit_eff}
\begin{figure}[h]
    \centering
\includegraphics[width=\textwidth]{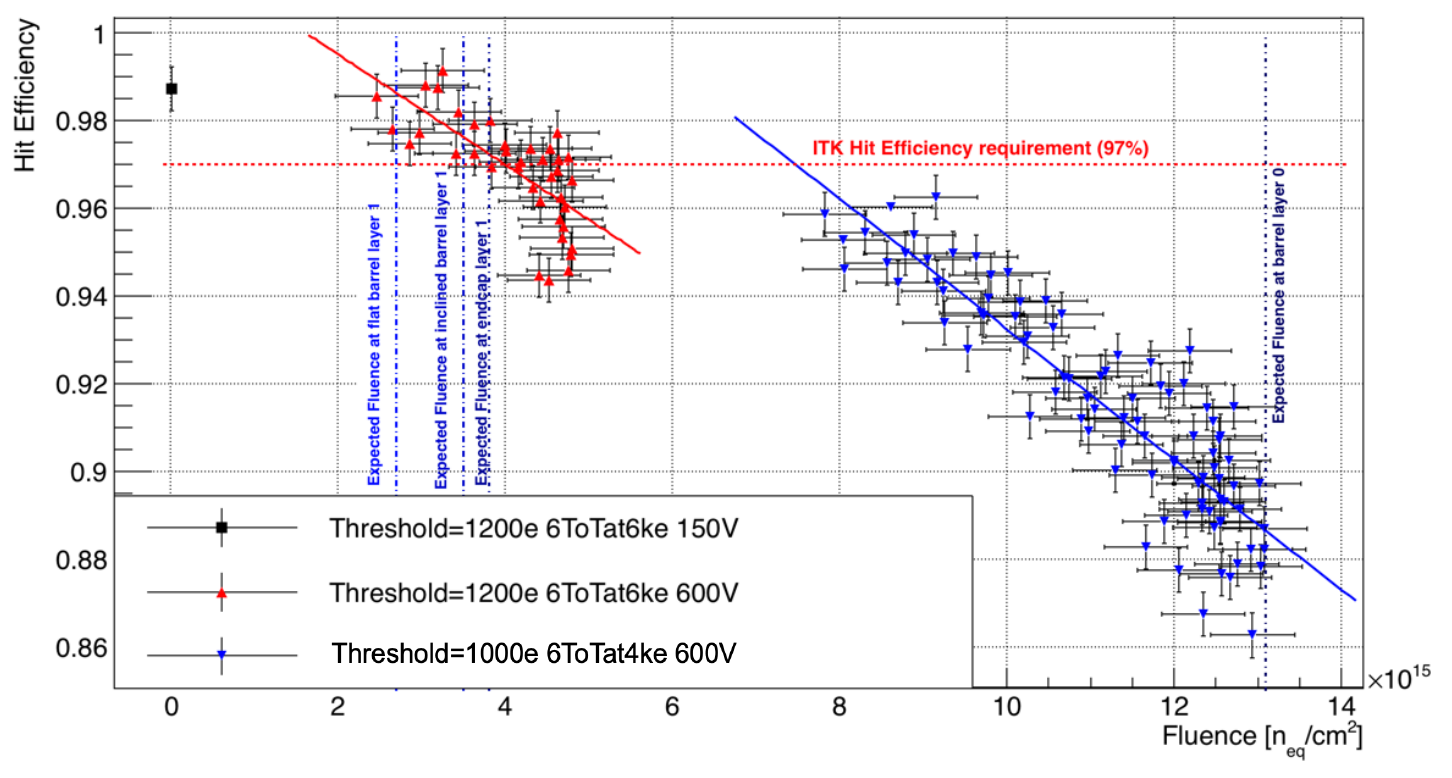} 

\caption{\label{fig:HitEff_vs_fluence}  Hit efficiency for a 130~$\mu$m thick sensor at various irradiation fluences. The blue triangles are for sensor irradiated at \fluence{1}{16} and the red ones at \fluence{3}{15}. The black square represents data for a thin un irradiated sensor. Threshold, ToT calibration and bias  voltage are indicated.}
\end{figure}

\noindent The hit efficiency $\epsilon$, defined as the fraction of reconstructed tracks crossing a module that have an 
associated hit in that module, was studied as a function of the irradiation fluence $\Phi$  using data collected at DESY 
testbeam facility, and contrasted with a result for an un-irradiated dectector. The results are presented in Figure~\ref{fig:HitEff_vs_fluence}.

The  efficiency versus fluence measurements of the two irradiation steps were fitted with a straight line. 
The horizontal uncertainty bars account for the uncertainty on the fluence peak position and on the modelling of the irradiation profiles, as 
explained in Section~\ref{sec:corrections} and~\ref{sec:OtherSyst}.
The vertical error bars are the combination of the systematic uncertainties arising from the selection criteria variations (0.4\%) and from 
the statistical fluctuations (0.25\%). For the statistical part, for all fluence points at least 5000 tracks were considered, hence the statistical 
(binomial) error is less than 0.25\%. The horizontal red dotted line represents the hit efficiency requirements of the ITk 
(97\%)~\cite{ATLASITkPixelTDR}. 
The four vertical blue dotted lines correspond the limit fluence expected at the end of lifetime of four different layers of ITk.
From lower to higher fluences one can find~\cite{ATLASITkPixelTDR}:
\begin{itemize}
\item the fluence expected (\fluence{2.7}{15}) at the layer 1 (second layer from the beam pipe) in the central (barrel) flat part;
\item the fluence expected (\fluence{3.5}{15}) at the layer 1 in the inclined part of the barrel;
\item the fluence expected (\fluence{3.8}{15})  at the layer 1 in the endcap part;
\item the fluence expected (\fluence{1.3}{16}) at the layer 0 (closest layer to the beam pipe) in the flat barrel part.
\end{itemize}

 Table~\ref{tab:eff_fluences} presents the expected efficiency for the various fluences, obtained from the crossing point of the fit and the fluence lines. The fluence ($\sim$\fl{7.45}{15}) at which the hit efficiency is 97\% is reported too.
\begin{table}
\caption{\label{tab:eff_fluences}Extrapolated efficiency for ITk benchmarks fluences for a 130~$\mu$m thick sensor.}
\centering
\begin{tabular}{|c|c|c|c|c|c|}
\hline
Fluence (\fluence{1}{15}) & 2.7 & 3.5 & 3.8 & 7.45 & 13.1 \\
\hline
Threshold (electrons) & 1200 & 1200 & 1200 & 1000 & 1000 \\
\hline
ToT tuning (ToT corresponding to ke-) & 6 at 6 & 6 at 6 & 6 at 6 & 6 at 4 & 6 at 4 \\
\hline 
Extrapolated Hit Efficiency (\%) & 98.6 & 97.6 & 97.2 & 97.0 & 88.6 \\
\hline
\end{tabular}
\end{table}

The measured efficiency obtained for the fluences of the various layer 1 parts are all above the 97\% requirement. 
A lower threshold and a better tuning could 
certainly help to reach higher values in terms of efficiency. For example, the prediction from the 1000e threshold data, assuming a linear 
dependency shows that the crossing between the ITk requirement line and the extrapolated values happen around  \fluence{7}{15}. As a reminder, pixels detectors in 3D technology~\cite{PARKER1997328} 
 are the baseline for the innermost layer of the ITk Pixels detector; those pixels modules will be replaced at half-life of 
 the HL-LHC, after  an integrated luminosity of about 2000~fb$^{-1}$.

\subsection{Pixel Resolved Hit Efficiency}

\begin{figure}[h]
\centering 
\includegraphics[width=\textwidth]{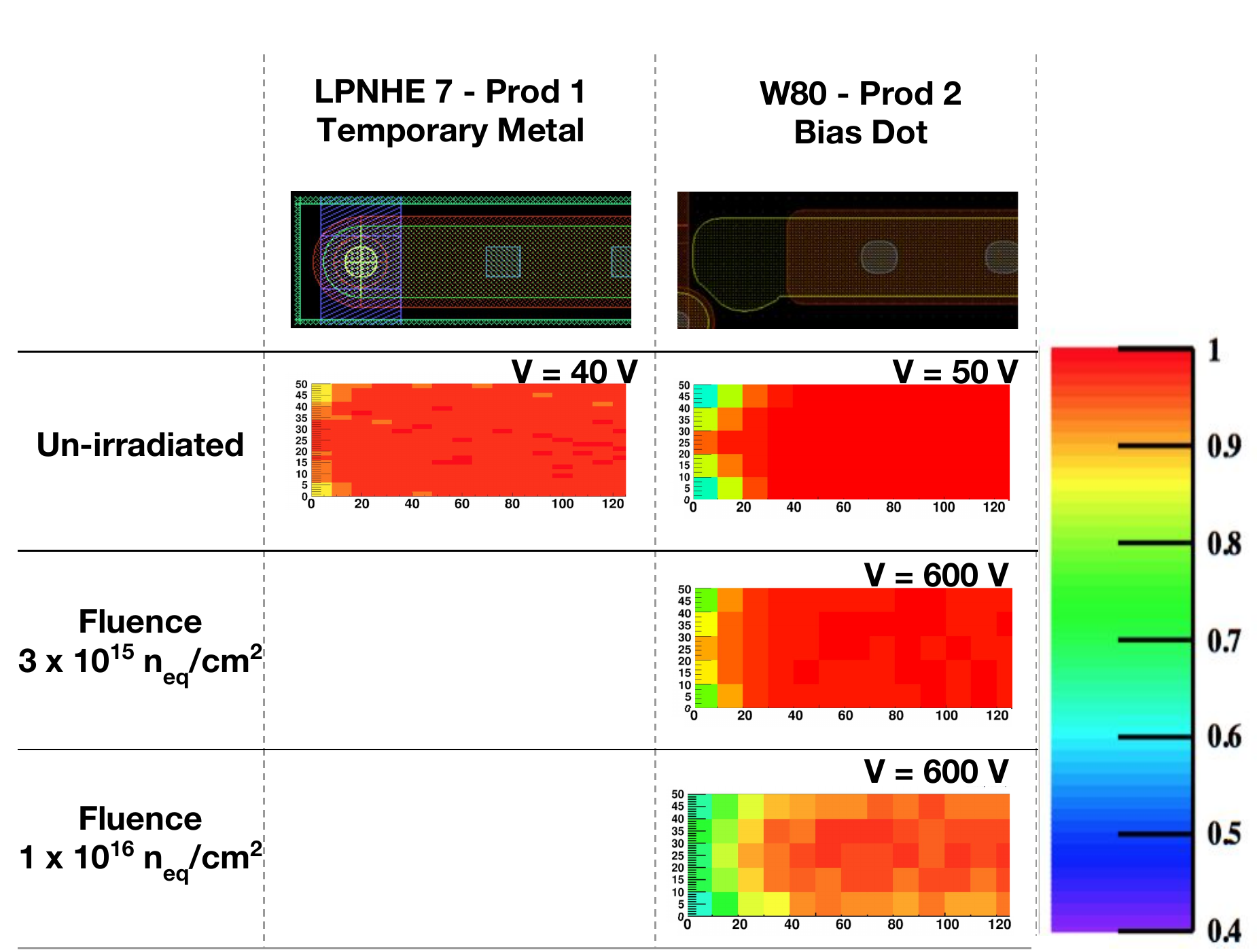} 
\caption{\label{fig:inpix_eff} Pixel resolved hit efficiency for two pixel sensors: left one from~\cite{1748-0221-12-05-P05006}, right one is 
W80. 
Results are presented for 3 different fluences for the W80; the corresponding sensor bias voltage is indicated. A scheme of the pixel is presented in the top part of the figure. See   Section~\ref{sec:sensors} and in particular Figure~\ref{fig:biasdot} and for details about the pixel cells parts.}
\end{figure}

\noindent In Figure~\ref{fig:inpix_eff} the hit efficiency for two different pixel detectors in several conditions is reported after having folded all cells in the 
matrix into one - the so-called \textit{pixel resolved hit efficiency}\footnote{This analysis is possible thanks to the good 
pointing resolution of the tracks telescope, even with the low energy beam of DESY.}.
 Data are presented for three different fluences; unirradiated, and irradiated with a fluence of  \fluence{3}{15} and of 
 \fluence{1}{16}. 
As indicated in Section~\ref{sec:intro} the W80 pixel sensor featured permanent biasing structure exploiting the 
 \textit{punch through} mechanism.  The punch through structure clearly degrades the performance at the corner of the sensor cell. This 
 is already evident before irradiation, especially when the pixel resolved hit efficiency is to compared to the case where 
 no permanent biasing structures are present (left column of the Figure; data from Ref.~\cite{1748-0221-12-05-P05006}).




\subsection{Charge Collection Performance}
\noindent Charge collection performance was studied as a function of fluence and bias voltage for W80 module.

\begin{figure}[h]
    \centering
\includegraphics[width=\textwidth]{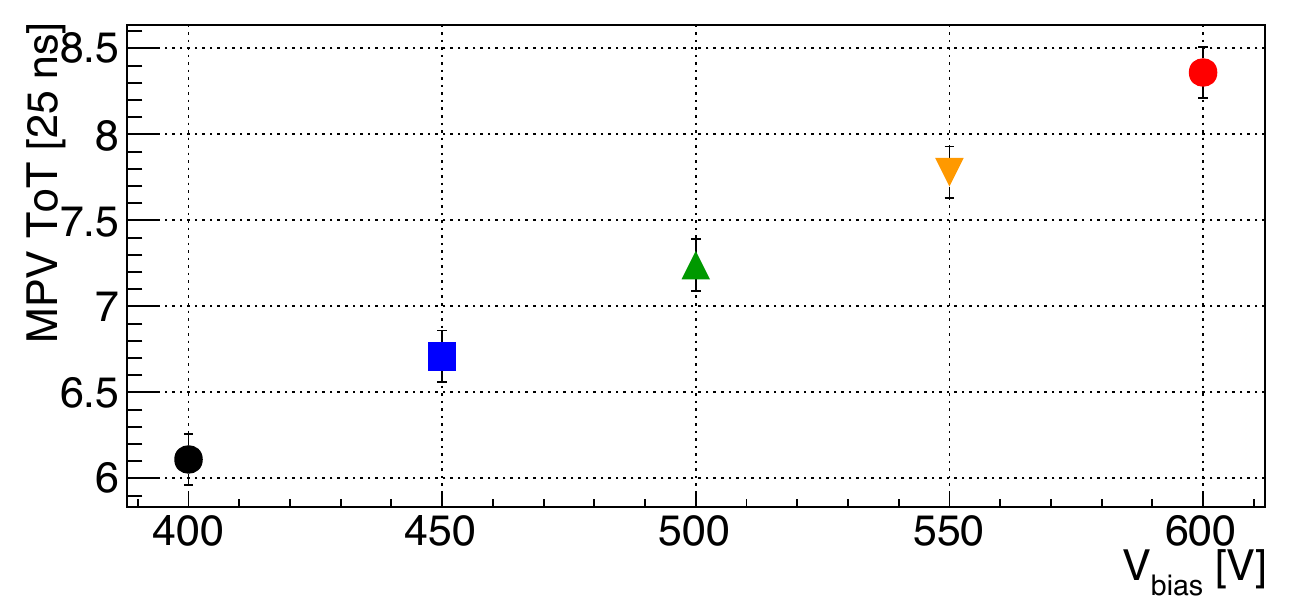}
\caption{\label{fig:MPV_vs_Bias} ToT distribution for thin sensor irradiated at \fluence{1}{16} (from~\cite{Ducourthial:2017kfw}).}
\end{figure}

\paragraph{Charge Collection vs Bias Voltage}
\noindent The analysis started looking at cluster ToT evolution with bias voltage for the W80 module after the second step of irradiation.
Data were taken at CERN and the level of beam collimation did not allow us to do the detailed fluence analysis proposed for data taken at 
DESY. Hence the fluence is averaged over the illuminated area. The tested configurations had a threshold of 850 electrons and the ToT to 
charge calibration was 8 ToT for 4000 electrons. All the distributions were fitted with a Landau distribution 
convoluted with a Gaussian resolution function allowing the 
determination of the Most Probable Value (MPV)~\cite{Ducourthial:2017kfw}.
In Figure~\ref{fig:MPV_vs_Bias}  
it can be seen that at high fluences the cluster ToT MPV increases linearly with the bias voltage in the range between 
400~V and 600~V.
At 400 V the MPV is 73\% of the one at 600 V.  At 600V the ToT MPV is $\simeq 8.5$, which corresponds  to a  charge 
slightly higher than 4000 electrons.
 By comparison with unirradiated sensor, this means that the collection efficiency is roughly reduced by a factor of 2 after an irradiation at 
 \fluence{1}{16}.

\paragraph{Charge Collection Efficiency vs Fluence}
Using the data collected at DESY - thanks to the wide beam spot - it was possible to investigate charge collection all over the 
pixel matrix and hence extract the charge collection performance as a function of the irradiation fluence. ToT to Charge calibrations  were 
performed using STControl~\cite{USBpix} software.

\begin{figure}[h]
    \centering
\includegraphics[width=\textwidth]{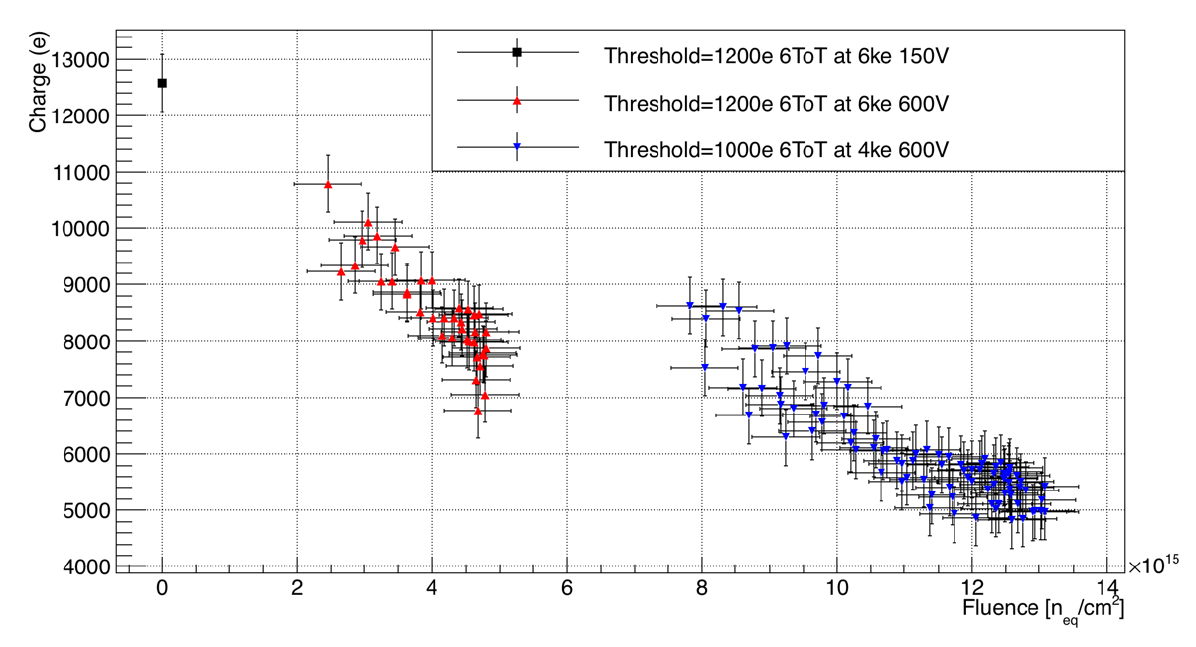}
\caption{\label{fig:Charge_vs_fluence} Collected charge vs Fluence for two sets of tuning: Threshold=1200e, 6ToT at 6ke (150~V bias voltage, un-irradiated and 600~V bias voltage, irradiated) and Threshold=1000e, 6ToT at 4ke, 600~V 
bias voltage.}
\end{figure}

The collected charge for the irradiated W80 module and for the un-irradiated W30 module is plotted in 
Figure~\ref{fig:Charge_vs_fluence}.


This plot compiles results from 3 testbeams where the W30 sensor was tested un-irradiated, biased at 150 V and with a threshold of 
1200e and a ToT configuration of 6ToT corresponding to 6000 electrons (black square on the plot), and W80 sensor was tested 
after the first irradiation step (red triangles) and after the second irradiation step (blue triangles). Before irradiation, 
the mean 
collected charge is 12500 electrons which is quite close to what is expected for a 100~$\mu$m thick sensor. The decrease with fluence of 
collected charge is steeper at lower fluences (red markers) than at higher fluences (blue markers). This is probably due to a threshold 
tuning of poorer quality of the former with respect to the latter. It can also be seen that the collected charge at the highest tested 
fluence (\fluence{1.3}{16}) is greater than 4000~e at 600~V.

From Figure~\ref{fig:Charge_vs_fluence}  the charge collection efficiency (CCE) was derived. 
The charge value before irradiation was obtained 
from the W30 sensor (100~$\mu$m thick). For the other fluences, the charge reported in Figure~\ref{fig:Charge_vs_fluence} was obtained 
from the W80 sensor (130~$\mu$m thick). Consequently the normalisation for the irradiated module was 1.3 times higher than for the 
un-irradiated one. 

\begin{figure}[h]
    \centering
\includegraphics[width=\textwidth]{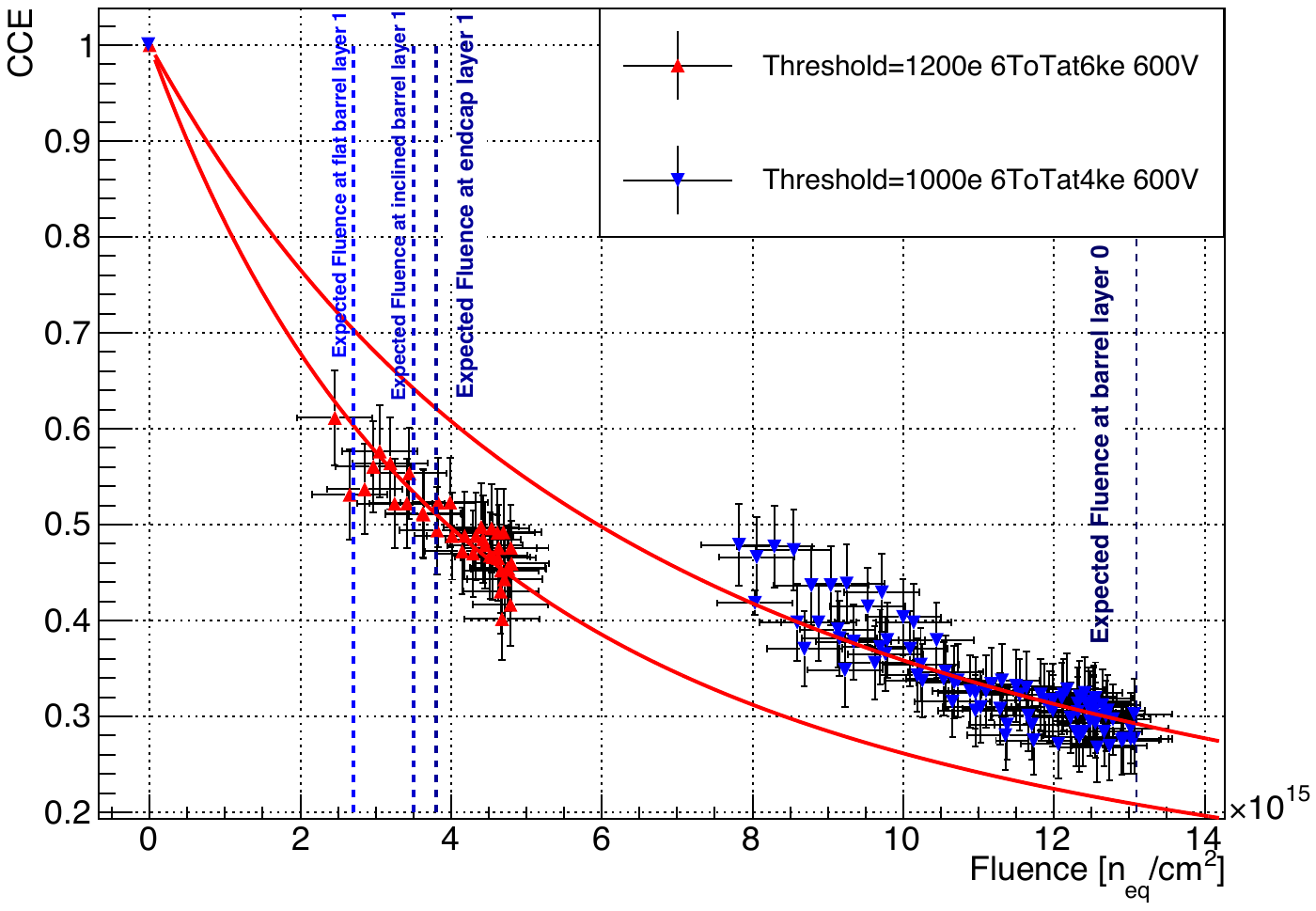}
\caption{\label{fig:cce_vs_fluence}Charge collection efficiency (CCE) measurement for two sets of tuning: Threshold=1200e, 
6ToT at 6000e and Threshold=1000e, 6ToT at 4000e. Expected end-of-lifetime fluences for ITk are also indicated. Data 
were collected at 600~V bias voltage.
Red line is a fit on data.}
\end{figure}

Figure~\ref{fig:cce_vs_fluence} shows the CCE of W80 over a typical ITk like fluence range. The distribution is fitted with the Hecht 
function (\cite{Hecht}; see also Appendix~A in~\cite{bomben_HDR}). From these fits  the effective trapping constant $\beta$ can be
 extracted; as a first approximation $\beta$ was assumed to be equal for holes and electrons. 
For the intermediate fluence dataset (red triangles) the following value for the trapping constant was extracted: 
$\beta = 5.3\pm 0.2\times 10^{-16}$~cm$^2$/ns; 
for the higher fluence dataset (blue triangles) a value of $\beta = 3.5\pm 0.1\times 10^{-16}$~cm$^2$/ns was fitted. The two values are of 
the same order of magnitude, their differences  come from the different tuning configurations, from a different annealing time or from 
the various approximations used. The fitted values reported here are somewhat smaller than those listed in Ref.~\cite{Trapping} 
(4-6 $\times10^{-16}$~cm$^2$/ns); 
this can 
be in part explained by the fact that the Hecht formula used to fit data assumes a mono-dimensional detector (\textit{i.e.} a pad detector 
whose 
sides are much larger than its thickness); the limitations of FE-I4 chip have also an impact. Hence the $\beta$ values obtained are to be considered as lower limits for the true ones.

The 4 vertical blue dotted lines represents the expected fluences at the end of lifetime of 4 different layers of 
ITk~\cite{ATLASITkPixelTDR} (see list in Section~\ref{sec:hit_eff}). The Table~\ref{tab:cce} compiles the values of the intersections of the
 2 fits with the 4 fluence lines.
 
 \begin{table}[htbp]
\caption{\label{tab:cce}Charge collection efficiency for two tunings over the ITk fluence range.}
\label{Tab:SRNRValues}
\begin{center}
\begin{tabular}{|c|c|c|c|c|}
\hline
\multirow{1}{2.5cm}{Calibration}& \multicolumn{4}{p{5cm}|}{\centering Fluences (\fluence{1}{15})} \\
\cline{2-5} & \multicolumn{1}{c|}{2.7} & \multicolumn{1}{c|}{3.5} & \multicolumn{1}{c|}{3.8} & \multicolumn{1}{c|}{13.1} \\ \hline
Thr.=1200e, 6ToT at 6000e, CCE = & 61\% & 53\% & 51\% & 21\% \\
\hline
Thr.=1000e, 6ToT at 4000e, CCE = & 71\% & 64\% & 62\% & 29\% \\
\hline
\end{tabular}
\end{center}
\end{table}
For the 1000 electrons threshold and 
the three fluences corresponding to the accumulated dose at layer 1, the charge collection efficiency is higher than 60\%. 
At \fluence{1.3}{16}, the fluence expected at Layer 0 in the flat section after 2000~fb$^{-1}$, the charge collection efficiency is lower than 
30\%.

\section{Conclusions}
\label{sec:conclusions}
\noindent Planar pixels sensors produced at FBK Trento by LPNHE and INFN 
were tested in beam before and after irradiation to fluences comparable to those 
expected at the end of the HL-LHC. Results indicate that the detectors meet all the specifications of the ATLAS ITk for all but the 
very innermost pixel layers; in particular hit efficiency is as high as 97\% for fluences up to to \fluence{7}{15}. Collected 
data allowed also the estimation of the trapping constant, even if the accuracy is limited by pixel geometry modelling and the 
 FE-I4 chip performance. 
A new pixel production at FBK is completed and detector prototypes are being measured; the available sensors are 100~$\mu$m 
thick and compatible with the new readout chip for the ITk pixels modules (RD53A~\cite{RD53}). 

\section*{Acknowledgments}
\noindent Planar pixel production presented in this paper  are supported by the Italian National Institute for Nuclear Research (INFN), Projects 
ATLAS, CMS, RD-FASE2 (CSN1) and by the H2020 project AIDA-2020, GA no. 654168. Some of the measurements leading to these results 
have been performed at the Test Beam Facility at DESY Hamburg (Germany), a member of the Helmholtz Association (HGF). The 
authors want to thanks the CERN IRRAD team for helping with the irradiation of the sensors and everyone involved in the support of the 
testbeam activity in the CERN North Area Test Beam Facility for making some of the measurements presented here possible. 






\bibliographystyle{elsarticle-num}
\bibliography{biblio}{}

\begin{thebibliography}{10}
\expandafter\ifx\csname url\endcsname\relax
  \def\url#1{\texttt{#1}}\fi
\expandafter\ifx\csname urlprefix\endcsname\relax\def\urlprefix{URL }\fi
\expandafter\ifx\csname href\endcsname\relax
  \def\href#1#2{#2} \def\path#1{#1}\fi

\bibitem{HL_LHC}
\href{http://hilumilhc.web.cern.ch/about/hl-lhc-project}{{The HL-LHC project}}.
\newline\urlprefix\url{http://hilumilhc.web.cern.ch/about/hl-lhc-project}

\bibitem{AtlasDetector}
{ATLAS Collaboration}, {The ATLAS Experiment at the CERN Large Hadron
  Collider}, JINST 3 (2008) S08003.
\newblock \href {http://dx.doi.org/10.1088/1748-0221/3/08/S08003}
  {\path{doi:10.1088/1748-0221/3/08/S08003}}.

\bibitem{AtlasID1}
{ATLAS collaboration}, \href{http://cdsweb.cern.ch/record/331063}{Inner
  detector - technical design report, 1}, Tech. rep., CERN (1997).
\newline\urlprefix\url{http://cdsweb.cern.ch/record/331063}

\bibitem{AtlasID2}
{ATLAS collaboration}, \href{http://cdsweb.cern.ch/record/381263}{Inner
  detector - technical design report, 2}, Tech. rep., CERN (1997).
\newline\urlprefix\url{http://cdsweb.cern.ch/record/381263}

\bibitem{HL-LHC}
S.~McMahon, P.~Allport, H.~Hayward, B.~Di~Girolamo,
  \href{https://cds.cern.ch/record/1952548}{{Initial Design Report of the ITk:
  Initial Design Report of the ITk}}, Tech. Rep. ATL-COM-UPGRADE-2014-029,
  CERN, Geneva (Oct 2014).
\newline\urlprefix\url{https://cds.cern.ch/record/1952548}

\bibitem{ITkStripsTDR}
\href{https://cds.cern.ch/record/2257755}{{Technical Design Report for the
  ATLAS Inner Tracker Strip Detector}} (Apr 2017).
\newline\urlprefix\url{https://cds.cern.ch/record/2257755}

\bibitem{ATLASITkPixelTDR}
{ATLAS Collaboration}, \href{https://cds.cern.ch/record/2285585}{{Technical
  Design Report for the ATLAS Inner Tracker Pixel Detector}}, Tech. Rep.
  CERN-LHCC-2017-021. ATLAS-TDR-030, CERN, Geneva (Sep 2017).
\newline\urlprefix\url{https://cds.cern.ch/record/2285585}

\bibitem{IBLTDR}
M.~Capeans, G.~Darbo, K.~Einsweiller, M.~Elsing, T.~Flick, M.~Garcia-Sciveres,
  C.~Gemme, H.~Pernegger, O.~Rohne, R.~Vuillermet,
  \href{https://cds.cern.ch/record/1291633}{{ATLAS Insertable B-Layer Technical
  Design Report}}, Tech. Rep. CERN-LHCC-2010-013. ATLAS-TDR-19, {CERN} (Sep
  2010).
\newline\urlprefix\url{https://cds.cern.ch/record/1291633}

\bibitem{IBL_paper}
B.~Abbott, et~al., {Production and Integration of the ATLAS Insertable
  B-Layer}, JINST 13~(05) (2018) T05008.
\newblock \href {http://dx.doi.org/10.1088/1748-0221/13/05/T05008}
  {\path{doi:10.1088/1748-0221/13/05/T05008}}.

\bibitem{KRAMBERGER2002297}
G.~Kramberger, V.~Cindro, I.~Mandi\v{c}, M.~Miku\v{z}, M.~Zavrtanik, Effective
  trapping time of electrons and holes in different silicon materials
  irradiated with neutrons, protons and pions, Nucl. Instr. and Meth. A 481~(1)
  (2002) 297 -- 305.
\newblock \href {http://dx.doi.org/10.1016/S0168-9002(01)01263-3}
  {\path{doi:10.1016/S0168-9002(01)01263-3}}.

\bibitem{DALLABETTA2016388}
G.-F.~D. Betta, M.~Boscardin, M.~Bomben, M.~Brianzi, G.~Calderini, G.~Darbo,
  R.~Dell'Orso, A.~Gaudiello, G.~Giacomini, R.~Mendicino, M.~Meschini,
  A.~Messineo, S.~Ronchin, D.~Sultan, N.~Zorzi, The infn-fbk ``phase-2'' r\&d
  program, Nucl. Instr. and Meth. A 824 (2016) 388 -- 391, frontier Detectors
  for Frontier Physics: Proceedings of the 13th Pisa Meeting on Advanced
  Detectors.
\newblock \href {http://dx.doi.org/10.1016/j.nima.2015.08.074}
  {\path{doi:10.1016/j.nima.2015.08.074}}.

\bibitem{FEI4}
M.~Garcia-Sciveres, et~al., {The FE-I4 pixel readout integrated circuit}, Nucl.
  Instrum. Meth. A636 (2011) S155--S159.
\newblock \href {http://dx.doi.org/10.1016/j.nima.2010.04.101}
  {\path{doi:10.1016/j.nima.2010.04.101}}.

\bibitem{RD53}
\href{https://rd53.web.cern.ch/RD53/}{{RD53 Collaboration}}.
\newline\urlprefix\url{https://rd53.web.cern.ch/RD53/}

\bibitem{PSI46}
H.~C. Kastli, M.~Barbero, W.~Erdmann, C.~Hormann, R.~Horisberger, D.~Kotlinski,
  B.~Meier, {Design and performance of the CMS pixel detector readout chip},
  Nucl. Instrum. Meth. A565 (2006) 188--194.
\newblock \href {http://dx.doi.org/10.1016/j.nima.2006.05.038}
  {\path{doi:10.1016/j.nima.2006.05.038}}.

\bibitem{1748-0221-12-05-P05006}
M.~Bomben, A.~Ducourthial, et~al., {Performance of active edge pixel sensors},
  JINST 12~(05) (2017) P05006.
\newblock \href {http://dx.doi.org/10.1088/1748-0221/12/05/P05006}
  {\path{doi:10.1088/1748-0221/12/05/P05006}}.

\bibitem{bib:metal}
E.~Vianello, A.~Bagolini, P.~Bellutti, M.~Boscardin, G.-F. Betta, G.~Giacomini,
  C.~Piemonte, M.~Povoli, N.~Zorzi, Optimization of double-side 3d detector
  technology for first productions at fbk, in: Nuclear Science Symposium and
  Medical Imaging Conference (NSS/MIC), 2011 IEEE, 2011, pp. 523--528.
\newblock \href {http://dx.doi.org/10.1109/NSSMIC.2011.6154102}
  {\path{doi:10.1109/NSSMIC.2011.6154102}}.

\bibitem{Stefano}
S.~Terzo, \href{https://mediatum.ub.tum.de/doc/1276352/1276352.pdf}{Development
  of radiation hard pixel modules employing planar n-in-p silicon sensors with
  active edges for the atlas detector at hl-lhc}, Ph.D. thesis, Technische
  Universitat Munchen, Max-Planck-Institut fur Physik (2015).
\newline\urlprefix\url{https://mediatum.ub.tum.de/doc/1276352/1276352.pdf}

\bibitem{UNNO201372}
Y.~Unno, et~al., Development of novel n+-in-p silicon planar pixel sensors for
  hl-lhc, Nucl. Instrum. Meth. A699 (2013) 72 -- 77.
\newblock \href {http://dx.doi.org/10.1016/j.nima.2012.04.061}
  {\path{doi:10.1016/j.nima.2012.04.061}}.

\bibitem{moll-thesis}
M.~Moll,
  \href{http://www-library.desy.de/cgi-bin/showprep.pl?desy-thesis99-040}{{Radiation
  damage in silicon particle detectors: Microscopic defects and macroscopic
  properties}}, Ph.D. thesis, Hamburg U. (1999).
\newline\urlprefix\url{http://www-library.desy.de/cgi-bin/showprep.pl?desy-thesis99-040}

\bibitem{Chilingarov_tscale}
A.~Chilingarov, Temperature dependence of the current generated in si bulk,
  Journal of Instrumentation 8~(10) (2013) P10003.
\newblock \href {http://dx.doi.org/10.1088/1748-0221/8/10/P10003}
  {\path{doi:10.1088/1748-0221/8/10/P10003}}.

\bibitem{Savic:2016xwn}
N.~Savic, J.~Beyer, A.~Macchiolo, R.~Nisius, {Investigation of thin n-in-p
  planar pixel modules for the ATLAS upgrade}, JINST 11~(12) (2016) C12008.
\newblock \href {http://dx.doi.org/10.1088/1748-0221/11/12/C12008}
  {\path{doi:10.1088/1748-0221/11/12/C12008}}.

\bibitem{Jansen2016}
S.~Jansen, H. Spannagel et~al, Performance of the eudet-type beam telescopes,
  EPJ Techniques and Instrumentation 3~(1) (2016) 7.
\newblock \href {http://dx.doi.org/10.1140/epjti/s40485-016-0033-2}
  {\path{doi:10.1140/epjti/s40485-016-0033-2}}.

\bibitem{1748-0221-7-10-P10028}
J.~Weingarten, S.~Altenheiner, M.~Beimforde, M.~Benoit, M.~Bomben, et~al.,
  {Planar Pixel Sensors for the ATLAS Upgrade: Beam Tests results}, JINST 7
  (2012) P10028.
\newblock \href {http://dx.doi.org/10.1088/1748-0221/7/10/P10028}
  {\path{doi:10.1088/1748-0221/7/10/P10028}}.

\bibitem{mimosa26}
C.~Hu-Guo, et~al., First reticule size \{MAPS\} with digital output and
  integrated zero suppression for the eudet-jra1 beam telescope, Nucl. Instr.
  and Meth. A 623~(1) (2010) 480 -- 482, 1st International Conference on
  Technology and Instrumentation in Particle Physics.
\newblock \href {http://dx.doi.org/10.1016/j.nima.2010.03.043}
  {\path{doi:10.1016/j.nima.2010.03.043}}.

\bibitem{eutelescope}
\href{http://eutelescope.web.cern.ch/}{http://eutelescope.web.cern.ch/}.
\newline\urlprefix\url{http://eutelescope.web.cern.ch/}

\bibitem{tbmon2}
\href{https://bitbucket.org/TBmon2/tbmon2/overview}{https://bitbucket.org/TBmon2/tbmon2/overview}.
\newline\urlprefix\url{https://bitbucket.org/TBmon2/tbmon2/overview}

\bibitem{PARKER1997328}
S.~Parker, C.~Kenney, J.~Segal, 3d - a proposed new architecture for
  solid-state radiation detectors, Nucl. Instr. and Meth. A 395~(3) (1997) 328
  -- 343.
\newblock \href {http://dx.doi.org/10.1016/S0168-9002(97)00694-3}
  {\path{doi:10.1016/S0168-9002(97)00694-3}}.

\bibitem{Ducourthial:2017kfw}
A.~Ducourthial, et~al., {Thin and edgeless sensors for ATLAS pixel detector
  upgrade}, JINST 12~(12) (2017) C12038.
\newblock \href {http://dx.doi.org/10.1088/1748-0221/12/12/C12038}
  {\path{doi:10.1088/1748-0221/12/12/C12038}}.

\bibitem{USBpix}
M.~Backhaus, et~al., Development of a versatile and modular test system for
  {ATLAS} hybrid pixel detectors, Nucl. Instr. Meth. A 650~(1) (2011) 37 -- 40,
  international Workshop on Semiconductor Pixel Detectors for Particles and
  Imaging 2010.
\newblock \href {http://dx.doi.org/10.1016/j.nima.2010.12.087}
  {\path{doi:10.1016/j.nima.2010.12.087}}.

\bibitem{Hecht}
K.~Hecht, {Zum Mechanismus des lichtelektrischen Prim\"{a}rstromes in
  isolierenden Kristallen}, Zeit. Physik. (1932) 77:235.

\bibitem{bomben_HDR}
M.~Bomben, \href{https://tel.archives-ouvertes.fr/tel-01824535}{{Silicon
  Trackers for High Luminosity Colliders}}, Habilitation {\`a} diriger des
  recherches, {Universit{\'e} Paris Diderot (Paris 7) Sorbonne Paris Cit{\'e}}
  (Mar. 2018).
\newline\urlprefix\url{https://tel.archives-ouvertes.fr/tel-01824535}

\bibitem{Trapping}
G.~Kramberger, et~al., Effective trapping time of electrons and holes in
  different silicon materials irradiated with neutrons, protons and pions,
  Nucl. Instr. and Meth. A 481~(1-3) (2002) 297 -- 305.
\newblock \href {http://dx.doi.org/10.1016/S0168-9002(01)01263-3}
  {\path{doi:10.1016/S0168-9002(01)01263-3}}.

\end{thebibliography}







\end{document}